\documentclass[a4paper]{article}

\usepackage{risorse/pacchetto}

\begin{document}
\begin{titlepage}
	\centering
	\includegraphics[width=0.2\textwidth]{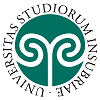}\par\vspace{1cm}
	{\scshape\LARGE University of Insubria \par}
	\vspace{.35cm}
	{\scshape\Large Department of Science and High Technology \par}
	\vspace{1cm}
	{\scshape\large Minor in Quantum Technologies\par}
	\vspace{2cm}
	{\huge\bfseries A four-state quantum communication protocol with mesoscopic twin beams\par}
	\vspace{0.35cm}
	{\Large\bfseries Channel characterization, state discrimination, and detection of intercept-resend attacks\par}
	\vspace{2cm}
	{\Large\itshape Stefano Carsi\par}
	\vfill
	Supervisors\par
	Prof.ssa~Alessia~\textsc{Allevi}\par
	Prof.ssa~Maria~\textsc{Bondani}\par
	\vspace{.5cm}
	Lab Assistant\par
	Dr.~Alex~\textsc{Pozzoli}\par

	\vfill

	{\large Year: 2026\par}
\end{titlepage}

\tableofcontents
\pagebreak

\section{Introduction}

Twin-beam (TWB) states are pairs of optical beams generated by parametric
down-conversion. The two beams, called signal and idler, are strongly correlated
in the number of photons: a measurement on one arm predicts the photon number in
the other better than any classical source of light. In the mesoscopic regime,
where each beam carries from a few up to some tens of photons, these states can
be measured directly with photon-number-resolving (PNR) detectors such as silicon
photomultipliers (SiPMs), and they keep their nonclassical character even in the
presence of moderate losses. Because real detectors have a finite quantum
efficiency, the incident photon number $n$ is not directly accessible: it is a
theoretical quantity, and what we measure is the detected photon number $m$. All
quantities in this report are expressed in terms of $m$.

The strength of these correlations is quantified by the noise reduction factor
$R$, defined as the variance of the photon-number difference between the two arms
divided by the shot-noise level. Classical light gives $R \ge 1$, while $R < 1$
denotes sub-shot-noise correlations and is a sufficient criterion for
entanglement. The noise reduction factor is therefore both a measure of the
channel quality and a witness of its quantum nature.

These properties make TWB states a useful resource for secure communication. In
the scheme of Razzoli et al.~\cite{razzoli:hybrid-2025}, the information is
encoded in thermal noise signals of different mean photon number, superimposed on
one arm of the TWB and sent to the receiver. The receiver reads the bits from the
mean photon number $\langle m \rangle$, a classical quantity that PNR detectors
estimate accurately. Security comes from $R$: an eavesdropper who intercepts part
of the signal and resends uncorrelated light leaves the mean value unchanged but
increases $R$, so the attack becomes visible. This binary protocol is taken as the
reference.

In this work the scheme is extended from two to four symbols, encoding two bits per
transmission. The four states are obtained by combining two mean values and two
mode numbers of the noise signal, which place them at four distinct points in the
$(\langle m \rangle, R)$ plane. The mean value separates one bit, while $R$
separates the other and, at the same time, provides the security check.

The analysis follows the two measurement campaigns. The twin beam is first
characterized alone (photon-number statistics, noise reduction factor, and detection
efficiency) to fix the working point of the channel. The four-state protocol is
then built with the superimposed noise, and its performance is studied: the state
constellation with its $95\%$ confidence regions, the error probability as a
function of the sample size, a comparison of machine learning classifiers for
state discrimination, and the response of $R$ to an intercept-and-resend attack.

\section{Theory}
\label{sec:theory}

This section collects the physical states and the quantities used throughout the
report. They follow the framework of Razzoli et al.~\cite{razzoli:hybrid-2025}
and of the review on mesoscopic twin beams~\cite{allevi:multimode-2022}. All
formulas are written in terms of the detected photon number $m$, the only
quantity we can access.

\subsection{Photon-number statistics of light}
\label{sec:stat}

The natural basis for the photon number is the Fock basis $\{\ket{n}\}$, the
eigenstates of the number operator $N=a^{\dagger}a$, with $N\ket{n}=n\ket{n}$.
Here $a$ and $a^{\dagger}$ are the ladder operators of the field mode: $a$ lowers
the photon number by one, $a\ket{n}=\sqrt{n}\,\ket{n-1}$, which is why it is
called the annihilation operator, while $a^{\dagger}$ raises it,
$a^{\dagger}\ket{n}=\sqrt{n+1}\,\ket{n+1}$, the creation operator. Only Fock
states have a definite number of photons. Any other state is a superposition or a
statistical mixture of Fock states and carries its own photon-number distribution
$P(n)$: the spread is a property of the state, not of the measurement, which only
collapses the state onto one value of $n$, drawn with the probability $P(n)$
assigns to it. Detection acts on top of this, because a detector of finite
efficiency registers a smaller number $m$ (Sec.~\ref{sec:fano}). To keep the
notation light we still speak of \textit{$n$ incident photons}, meaning the $n$
energy quanta the state exchanges with the detector in a single shot. Two cases
are relevant here, coherent light and thermal light.

A coherent state $\ket{\alpha}$ is the eigenstate of the annihilation operator
with eigenvalue $\alpha\in\C$, that is \mbox{$a\ket{\alpha}=\alpha\ket{\alpha}$}.
It describes the light of an ideal single-mode laser, with a well-defined
amplitude and phase. In the Fock basis it expands as
\begin{equation}
	\label{eq:coherent-fock}
	\ket{\alpha} = e^{-|\alpha|^{2}/2}\sum_{n=0}^{\infty}
	               \frac{\alpha^{n}}{\sqrt{n!}}\,\ket{n}
\end{equation}
with mean incident photon number $\langle n\rangle=|\alpha|^{2}$. The photon
number is then Poisson distributed, a form that detection preserves, so for the
detected photons
\begin{equation}
	\label{eq:poisson}
	P(m) = e^{-\langle m\rangle}\,\frac{\langle m\rangle^{m}}{m!}
\end{equation}
for which the variance equals the mean, $\sigma^{2}(m)=\langle m\rangle$. It is
called coherent because the field keeps a stable, well-defined phase, the property
that lets light interfere. Among the quantum states it is the closest to a
classical wave.

Thermal, or chaotic, light comes from a source in thermal equilibrium. The same
statistics is produced by laser light scattered by a moving diffuser, when a
fixed region the size of a single speckle is collected: the speckle pattern
changes from shot to shot, so the collected intensity fluctuates, whereas a
frozen speckle would give a constant intensity (Sec.~\ref{sec:setup-noise}).
Unlike the coherent state it is a statistical mixture with no definite phase,
described by the density operator
\begin{equation}
	\label{eq:thermal-rho}
	\rho_{\mathrm{th}} = \sum_{n=0}^{\infty} P(n)\,\ket{n}\bra{n}
\end{equation}
diagonal in the Fock basis, where the weights $P(n)$ follow the Boltzmann
distribution over the energy levels of the mode. For a single mode this gives a
geometric distribution which, like the Poissonian case, keeps its form under
detection, so for the detected photons
\begin{equation}
	\label{eq:thermal}
	P(m) = \frac{\langle m\rangle^{m}}{(1+\langle m\rangle)^{m+1}}
\end{equation}
with variance $\sigma^{2}(m)=\langle m\rangle+\langle m\rangle^{2}$, larger than
the coherent case. When $\mu$ equally populated modes
are detected together the distribution generalizes to the multi-mode thermal
distribution (MMTD)
\begin{equation}
	\label{eq:mmtd}
	P^{\mu}(m) =
	\frac{(m+\mu-1)!}{m!\,(\mu-1)!}\;
	\frac{1}{\left(\dfrac{\langle m\rangle}{\mu}+1\right)^{\mu}
	         \left(\dfrac{\mu}{\langle m\rangle}+1\right)^{m}}
\end{equation}
with mean $\langle m\rangle$, variance $\sigma^{2}(m)=\langle m\rangle+\langle
m\rangle^{2}/\mu$, and effective number of modes $\mu$. The single mode $\mu=1$
recovers Eq.~\eqref{eq:thermal}, while for $\mu\to\infty$ the variance tends to
the Poissonian value and the distribution approaches Eq.~\eqref{eq:poisson}. When
$\mu$ is treated as a continuous fit parameter, the factorials are continued
through the Euler Gamma function, $x!=\Gamma(x+1)$; here $\Gamma$ denotes the Gamma
function, not the correlation coefficient $\Gamma$ of Eq.~\eqref{eq:gamma}.

A complementary descriptor of the statistics is the normalized second-order
correlation function at zero time delay
\begin{equation}
	\label{eq:g2}
	g^{(2)}(t=0) = \frac{\langle m(m-1)\rangle}{\langle m\rangle^{2}}
\end{equation}
which equals \num{1} for coherent light and $1+1/\mu$ for multi-mode thermal light,
that is \num{2} for a single thermal mode. It measures photon bunching: thermal light
is bunched ($g^{(2)}(t=0)>1$), while a coherent state shows no bunching.

\subsection{Twin-beam states}
\label{sec:twb}

A twin-beam (TWB) state is generated by spontaneous parametric down-conversion in
a nonlinear crystal, where a pump photon is converted into a signal and idler
photon pair by interacting with the vacuum fluctuations of the two fields. The two
photons of each pair obey energy and
momentum conservation, $\omega_p=\omega_s+\omega_i$ and
$\vec{k}_p=\vec{k}_s+\vec{k}_i$, the latter known as the phase-matching condition.
These laws correlate signal and idler in frequency and direction, in addition to
the photon number, so that selecting a region on one arm fixes its twin on the
other. For $\mu$ equally populated modes the state is
\begin{equation}
	\label{eq:twb}
	\ket{\Psi^{\mu}_{\mathrm{TWB}}} =
	\bigotimes_{k=1}^{\mu}\sqrt{1-\lambda^{2}}
	\sum_{\nu=0}^{\infty}\lambda^{\nu}\,\ket{\nu}_k\otimes\ket{\nu}_k
\end{equation}
with $\lambda^{2}=\langle n\rangle/(\mu+\langle n\rangle)$ and $\langle n\rangle$
the mean number of incident photons in each arm. Every term contains
$\ket{\nu}\otimes\ket{\nu}$, the same photon number $\nu$ in both arms, so the two
beams are perfectly correlated in the number of photons. Each arm taken alone has
multi-mode thermal statistics, Eq.~\eqref{eq:mmtd}, and looks like ordinary
thermal light, with the nonclassical features appearing only when the two arms
are compared.

\subsection{Shot noise and the Fano factor}
\label{sec:fano}

The fluctuations of a light beam are measured against the shot-noise level, the
variance of a beam with Poissonian statistics. This is the case of a coherent
state, for which the variance of the photon number equals its mean. Shot noise
sets the classical reference: no classical state can fluctuate below it, while
nonclassical light can. The comparison is made through the Fano factor
\begin{equation}
	\label{eq:fano}
	F(m) = \frac{\sigma^{2}(m)}{\langle m\rangle}
\end{equation}
which equals \num{1} for Poissonian light, is smaller than \num{1} for sub-Poissonian
(nonclassical) light, and larger than \num{1} for super-Poissonian light, as for
thermal and multi-mode thermal states. Real detectors register only a fraction of
the incident photons: with an overall quantum efficiency $\eta$ each photon is
detected independently with probability $\eta$, so the detected mean is $\langle
m\rangle=\eta\langle n\rangle$ and the detected and incident Fano factors are
linked by
\begin{equation}
	\label{eq:fano-eta}
	F(m) = \eta\,F(n) + (1-\eta)
\end{equation}
so the deviation from the Poissonian value is scaled by the efficiency.
Subtracting \num{1} from both sides of Eq.~\eqref{eq:fano-eta} gives
$F(m)-1=\eta\,[F(n)-1]$: the same factor that reduces the mean also reduces any
excess or deficit of noise, and a detector of very low efficiency returns
Poissonian statistics whatever the incident state. For multi-mode thermal light the relation
reads $F(m)=1+\langle m\rangle/\mu$, so the super-Poissonian excess shrinks
together with the detected mean, which is why the beams measured here have $F$
only slightly above \num{1} (Sec.~\ref{sec:twb-stat}). This loss of information is
the reason all quantities in this report are expressed in terms of the detected
photon number $m$.

\subsection{Noise reduction factor and photon-number correlation}
\label{sec:R-gamma}

The key nonclassical quantity is the noise reduction factor $R$, defined from the
variance of the photon-number difference between the two arms
\begin{equation}
	\label{eq:R}
	R = \frac{\sigma^{2}(m_1-m_2)}{\langle m_1\rangle + \langle m_2\rangle}
\end{equation}
The denominator is the shot-noise level, equal to the variance of the
photon-number difference expected for two uncorrelated beams with the same mean
values. Classical light gives $R\ge 1$, while $R<1$ proves sub-shot-noise
correlations between the arms and is a sufficient criterion for entanglement. A
twin beam with perfectly correlated arms would give $R=0$, because the two photon
numbers fluctuate together and their difference is constant. The linear
correlation between the arms is described by the photon-number correlation
coefficient
\begin{equation}
	\label{eq:gamma}
	\Gamma = \frac{\langle m_1 m_2\rangle - \langle m_1\rangle\langle m_2\rangle}
	              {\sqrt{\sigma^{2}(m_1)\,\sigma^{2}(m_2)}}
\end{equation}
which takes values in $[-1,1]$.

\subsection{Noise reduction factor with losses and noise}
\label{sec:R-model}

In a realistic channel the two arms are unbalanced and a thermal noise signal is
added to one of them. Writing $t\in[0,1]$ for the transmission efficiency that
quantifies the imbalance between the arms, $\langle m\rangle$ for the mean
detected photons of a TWB arm, and $\langle m_N\rangle$ and $\mu_N$ for the mean
value and number of modes of the noise, the noise reduction factor becomes
\begin{equation}
	\label{eq:R-model}
	\begin{aligned}
		R = 1
		&- \frac{2\eta t\,\langle m\rangle}{(1+t)\langle m\rangle+\langle m_N\rangle} \\
		&+ \frac{(1-t)^{2}\langle m\rangle^{2}}{\mu\big[(1+t)\langle m\rangle+\langle m_N\rangle\big]} \\
		&+ \frac{\langle m_N\rangle^{2}}{\mu_N\big[(1+t)\langle m\rangle+\langle m_N\rangle\big]}
	\end{aligned}
\end{equation}
where $\mu$ is the number of modes of the TWB. Setting $\langle m_N\rangle=0$
describes a channel with losses but no added noise, while $t=1$ describes two arms
of equal mean.
Equation~\eqref{eq:R-model} shows the two effects that raise $R$ toward and above
\num{1}: the imbalance ($t<1$) and the added noise, which is uncorrelated with the
TWB. The last term also makes $R$ grow when the noise has fewer modes (small
$\mu_N$), because a smaller number of modes means larger relative fluctuations.
Nonclassical correlations ($R<1$) survive as long as
\begin{equation}
	\label{eq:nonclass}
	\langle m_N\rangle <
	\sqrt{\mu_N\big[2\eta t\mu-(1-t)^{2}\langle m\rangle\big]\,
	\frac{\langle m\rangle}{\mu}}
\end{equation}

\section{Experimental setup}
\label{sec:setup}

\begin{figure}[htp]
	\centering
	\includegraphics[width=\textwidth]{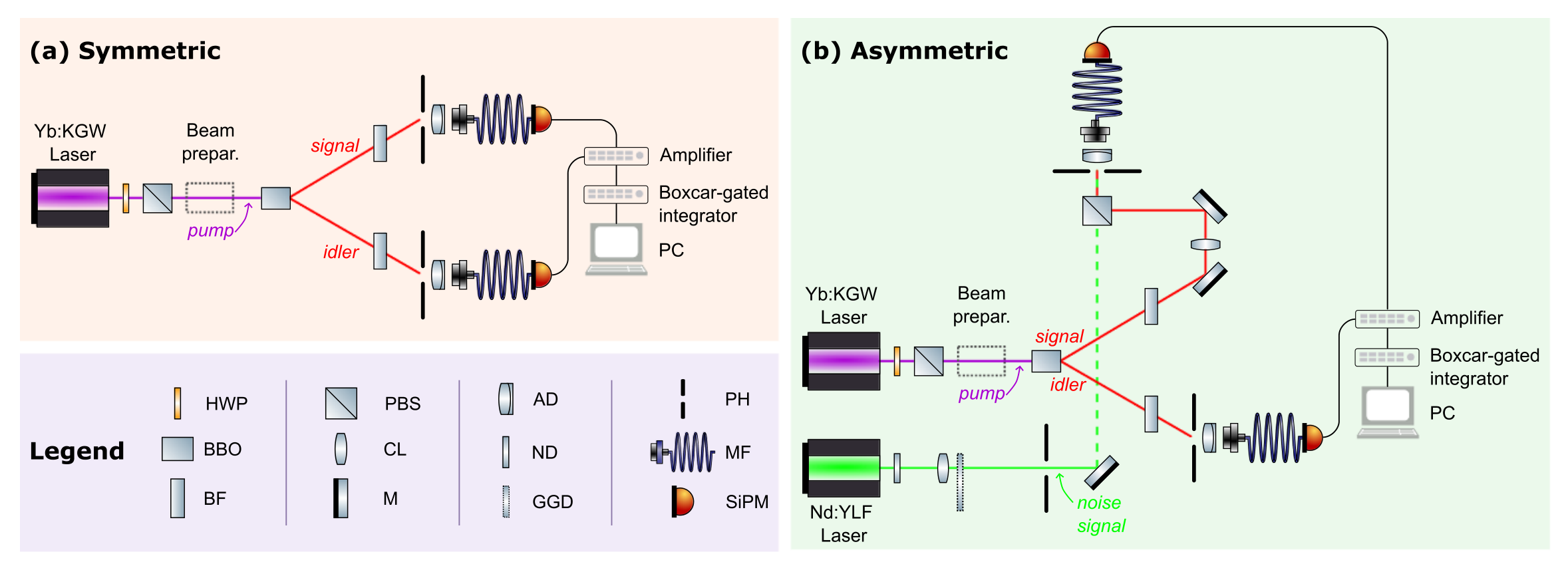}
	\caption{Experimental setup, reproduced from Razzoli et
	al.~\cite{razzoli:hybrid-2025}. (a) Symmetric
	configuration, used to characterize the twin beam. (b) Asymmetric
	configuration, used for the communication protocol: the noise signal (green,
	from a separate laser) is added to the signal arm. In the original work
	the two contributions were combined in post-processing (dashed line in the
	figure), while in the present measurements they are combined optically into
	the same collection fiber. The legend lists the optical components.}
	\label{fig:setup}
\end{figure}

The experimental apparatus is the one of Razzoli et
al.~\cite{razzoli:hybrid-2025}, except for the silicon photomultiplier model
(Sec.~\ref{sec:setup-detection}) and the laser that generates the noise
(Sec.~\ref{sec:setup-noise}); the reader is referred to that work for the
full description, and only the elements relevant to the present measurements are
summarized here. The setup has two parts: the
generation and detection of the twin beam, and the generation of the thermal noise
that encodes the symbols (Fig.~\ref{fig:setup}).
The two are produced by separate sources and combined optically on the signal
arm (Sec.~\ref{sec:setup-noise}).

\subsection{Twin-beam source}
\label{sec:setup-twb}

The twin beam is produced by parametric down-conversion in a
$\beta$-barium-borate (BBO) crystal, pumped by the third harmonic at
\SI{343}{\nano\meter} of a sub-picosecond Yb:KGW laser. Before the crystal the
pump pulses are tilted by a pair of prisms and a demagnifying telescope to achieve
group-velocity matching, so that the phase-matching condition is met near
frequency degeneracy at \SI{694}{\nano\meter}. The pump energy, and therefore the
mean photon number of the twin beam, is tuned with a half-wave plate followed by a
polarizing beam splitter at the laser output.

Signal and idler are filtered in the same way in the two arms, both spectrally,
with band-pass filters about \SI{20}{\nano\meter} wide centered at
\SI{692}{\nano\meter} and \SI{697}{\nano\meter}, and spatially, with an adjustable
pin-hole. Each arm is then focused by an achromatic doublet into a multi-mode
optical fiber with a \SI{1}{\milli\meter} core and sent to its detector. The two arms are not perfectly balanced:
in a real system the losses along the two paths are never identical, which gives a
transmission imbalance $t<1$ between signal and idler.

\subsection{Photon-number-resolving detection}
\label{sec:setup-detection}

Detection uses two silicon photomultipliers (SiPMs, Hamamatsu
S15639-1325PS~\cite{hamamatsu:s15639-datasheet}),
solid-state photon-number-resolving detectors made of \num{2120} cells with a
\SI{25}{\micro\meter} pitch over a $1.3\times1.1$ \si{\milli\meter\squared}
sensitive area, operated in the Geiger-Müller regime. Each cell fires when it is
hit by at least one photon, and the SiPM output is proportional to the number of
fired cells, which gives direct access to the photon number as long as the
probability of two photons on the same cell is small, a condition well satisfied
with thousands of cells and mesoscopic light. The detectors have a peak quantum
efficiency of about \SI{30}{\percent} at \SI{660}{\nano\meter} and are operated
at a bias voltage of \SI{54.9}{\volt}, \SI{10}{\volt} above breakdown, with a
gain of \num{1.3e6}. The typical dark count rate of \SI{700}{\kilo\hertz} is
negligible over the acquisition gate (below $10^{-2}$ counts per shot) and the
crosstalk probability is about \SI{4}{\percent}. The bias is supplied and the
output amplified by \SI{17}{\decibel} by a power supply and amplification unit
(CAEN SP5600), then integrated over a \SI{10}{\nano\second} gate by a boxcar
integrator and acquired. Alice measures the idler arm and Bob the signal arm.

\subsection{Noise signal generation}
\label{sec:setup-noise}

The thermal noise that encodes the symbols is produced by sending picosecond
pulses at \SI{515}{\nano\meter}, from a diode laser, onto a rotating ground-glass
disk. The scattered light forms a speckle pattern, and a small portion is selected
in the far field by an iris. The rotation of the disk is essential: it makes the
speckle pattern change from shot to shot, so the intensity collected at a fixed
point fluctuates over the acquisition with the statistics of thermal light, whereas
a static disk would give a frozen speckle and a constant intensity. This is the
standard way to obtain pseudo-thermal light, and the selected beam has the
multi-mode thermal statistics of Sec.~\ref{sec:stat}, with an effective number of
modes set by the size of the selected region, since a single speckle corresponds
to one mode and a larger region to more modes. The mean photon number is set
independently by a variable neutral-density filter placed between the laser and the
disk.
In this way the noise signals are prepared with the two mean values and the two mode
numbers needed for the four symbols. The noise and the twin beam are generated by
separate sources and then combined optically, by making the noise converge into
the same collection fiber as the signal arm, so that the two are physically added
before detection.

\subsection{From detector output to detected photons}
\label{sec:setup-processing}

Each acquisition produces a pulse-height spectrum in the arbitrary units of the
sampling ADC. Since the SiPM resolves the number of fired cells, the spectrum
shows a sequence of well-separated peaks, each corresponding to an integer number
of detected photons. The raw spectrum is converted into photon numbers by a program provided by the
laboratory, which automatically detects the peaks, locates the valley between
adjacent peaks at their midpoint, and places a threshold at each valley. Every event falling between two consecutive
thresholds is assigned to the same photon number, that of the peak they enclose.
In this way the arbitrary ADC units are mapped to integer detected-photon numbers,
on which all the following analysis is carried
out~\cite{bondani:self-consistent-2009}. For each configuration, \num{2e5} shots were acquired.

\section{Characterization of the twin beam}
\label{sec:twb-char}

In the first measurement campaign the twin beam is acquired alone, without any
noise signal, at five increasing pump intensities, with a mean detected photon
number between $\langle m\rangle\approx0.1$ and \num{6.4}. The two arms are well
balanced, with a transmission ratio $t\approx0.94$. The aim is to verify the
photon-number statistics of the source, to check the nonclassical correlations
between the arms, and to estimate the detection efficiency $\eta$ that enters the
model of the noise reduction factor. The measured quantities are collected in
Table~\ref{tab:twb}; their uncertainties are obtained by splitting each
$2\times10^{5}$-shot acquisition into eight disjoint batches and taking the
standard error of the batch values, while $\mu$ and $\eta$ come from the
corresponding fits. The number of modes is fitted separately on the two arms and
the table reports the average; since the two fits differ by more than their
errors, the uncertainty on $\mu$ combines in quadrature the fit error and the
half-difference between the arms, which dominates.

\begin{table}[htp]
	\centering
	\sisetup{separate-uncertainty=false}
	\caption{Twin-beam characterization. Columns 1 to 5 are the five acquisitions
	of increasing pump intensity. The first block reports the mean detected photons
	$\langle m_k\rangle$, the variances $\sigma_k^2$ and the arm ratio $t$ (smaller
	over larger arm mean, statistical error negligible); the second block the Fano
	factors $F_k$, the number of modes $\mu$ from the MMTD fit, the noise reduction
	factor $R$ and the correlation coefficient $\Gamma$. Uncertainties are in compact
	form, $\mathrm{value}(\text{error on the last digits})$. In columns 2 to 5 arm 2
	is always the smaller, while in column 1 the very low intensity inverts this and
	arm 1 becomes the smaller; the model uses the average of $t$ over columns 2 to
	5, where the arms are well balanced.}
	\label{tab:twb}
	\begin{tabular}{lccccc}
		\toprule
		 & 1 & 2 & 3 & 4 & 5 \\
		\midrule
		$\langle m_1\rangle$ & \num{0.09} & \num{1.03} & \num{2.05} & \num{4.56} & \num{6.40} \\
		$\langle m_2\rangle$ & \num{0.11} & \num{0.99} & \num{1.93} & \num{4.26} & \num{5.95} \\
		$\sigma_1^2$ & \num{0.09} & \num{1.09} & \num{2.19} & \num{4.79} & \num{6.73} \\
		$\sigma_2^2$ & \num{0.12} & \num{1.05} & \num{2.05} & \num{4.53} & \num{6.34} \\
		$t$ & \num{0.80} & \num{0.96} & \num{0.94} & \num{0.93} & \num{0.93} \\
		\midrule
		$F_1$ & \num{1.054(3)} & \num{1.057(3)} & \num{1.068(4)} & \num{1.051(4)} & \num{1.051(4)} \\
		$F_2$ & \num{1.049(5)} & \num{1.061(2)} & \num{1.061(1)} & \num{1.062(3)} & \num{1.066(5)} \\
		$\mu$ & \num{2.0(3)} & \num{17.6(9)} & \num{31.9(11)} & \num{83(13)} & \num{105(16)} \\
		$R$ & \num{0.9423(18)} & \num{0.9171(35)} & \num{0.9226(23)} & \num{0.9122(27)} & \num{0.9134(26)} \\
		$\Gamma$ & \num{0.104(2)} & \num{0.134(3)} & \num{0.133(1)} & \num{0.136(2)} & \num{0.137(1)} \\
		\bottomrule
	\end{tabular}
\end{table}

\subsection{Photon-number statistics}
\label{sec:twb-stat}

\begin{figure}[htp]
	\centering
	\includegraphics[width=.8\textwidth]{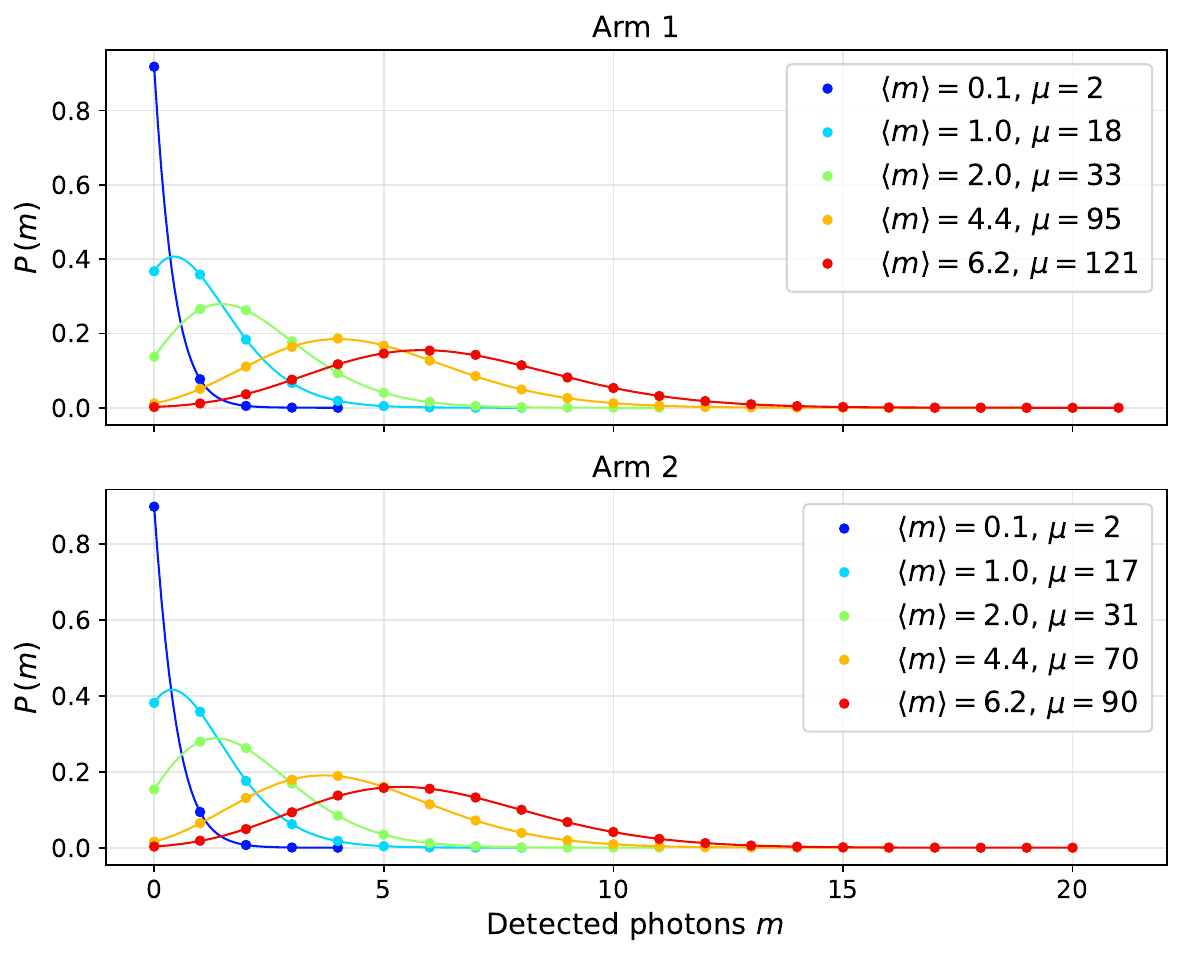}
	\caption{Detected-photon distributions of the two arms at the five pump
	intensities (markers) with the multi-mode thermal fits of
	Eq.~\eqref{eq:mmtd} (lines). The legend reports the mean photon number and the
	fitted number of modes.}
	\label{fig:twb_dist}
\end{figure}

Figure~\ref{fig:twb_dist} shows the detected-photon distributions of the two arms
with the multi-mode thermal fits of Eq.~\eqref{eq:mmtd}, where $\mu$ is a free
continuous parameter and the factorials are continued through the Gamma function,
$x!=\Gamma(x+1)$. The two arms follow the multi-mode thermal form at all five
intensities, with a Fano factor slightly above one ($F\approx1.05$,
Table~\ref{tab:twb}), as expected for thermal light.
The fitted number of modes $\mu$ grows with the intensity, from about \num{2} to
\num{100}. The low detection efficiency brings the statistics close to Poissonian,
with a Fano factor that stays nearly constant across the dataset, and since
$F=1+\langle m\rangle/\mu$ a constant $F$ means a $\mu$ proportional to the mean.

\subsection{Nonclassical correlations and detection efficiency}
\label{sec:twb-eta}

The noise reduction factor is below one for every intensity, with $R$ between
\num{0.91} and \num{0.94} (Table~\ref{tab:twb}), which proves the presence of
sub-shot-noise correlations between the two arms. The photon-number correlation
coefficient is $\Gamma\approx0.13$, stable over the four higher intensities,
while at the lowest one it drops to \num{0.104}. For an ideal
twin beam detected with unit efficiency the two arms would carry the same photon
number, giving $\Gamma=1$. The small measured value is a consequence of the finite
detection efficiency: we access only the detected photons, whose mean is
$\langle m\rangle=\eta\langle n\rangle$, and the
random independent losses on the two arms degrade the otherwise perfect
correlation. For a twin beam measured with efficiency $\eta$ on both arms the
coefficient becomes $\Gamma=(\eta+\langle m\rangle/\mu)/(1+\langle m\rangle/\mu)$;
since $\langle m\rangle/\mu=F-1\approx0.06$ is nearly constant across the dataset,
this predicts a nearly constant $\Gamma\approx0.14$ for all five acquisitions,
slightly above the measured plateau. The drop of the lowest-intensity point is not
reproduced because that expression assumes a pure twin beam with losses and no
background: at $\langle m\rangle\approx0.1$ the beam is weak enough for an
uncorrelated background to matter, since the dark counts alone are already several
percent of the signal (Sec.~\ref{sec:setup-detection}), and counts uncorrelated
between the arms lower $\Gamma$ without appearing in the model.

\begin{figure}[htp]
	\centering
	\includegraphics[width=\textwidth]{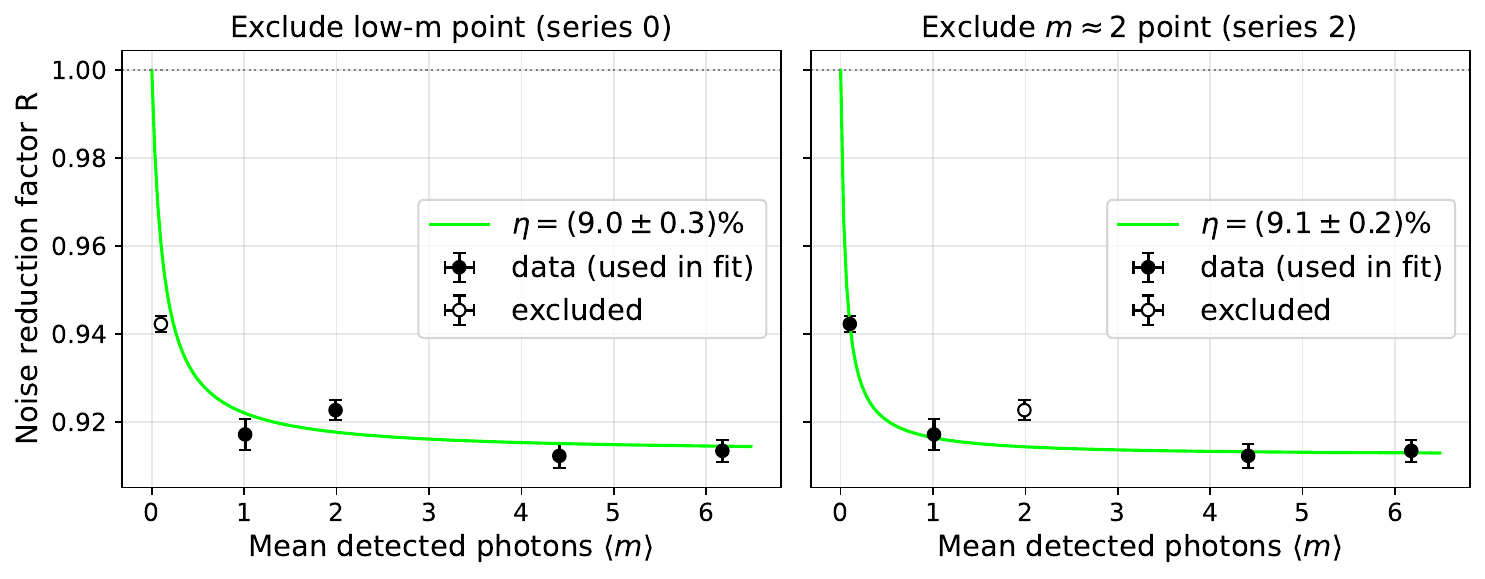}
	\caption{Noise reduction factor as a function of the mean detected photons,
	with the fit of the loss model. Left: the low-intensity point is excluded, and
	the curve extrapolated below the fitted range does not describe it. Right: the
	$\langle m\rangle\approx2$ point is excluded instead,
	keeping the low-intensity point. Both exclusions give the same efficiency within
	the uncertainty. The third fit discussed in the text, which keeps all five
	points, is not shown.}
	\label{fig:twb_R}
\end{figure}

The detection efficiency is estimated by fitting $R$ as a function of
$\langle m\rangle$ with Eq.~\eqref{eq:R-model}. No thermal noise is injected in
this campaign, so the last term is dropped, which is its Poissonian limit
$\mu_N\to\infty$, and $\langle m_N\rangle$ is kept as a free parameter that
accounts for a residual uncorrelated background. For balanced arms and
negligible background the model reduces to
$R\simeq1-2\eta t/(1+t)$, almost independent of $\langle m\rangle$, consistent
with the flat trend of the data. The lowest-intensity point
($\langle m\rangle\approx0.1$) is first excluded: there the counts include a
background that carries no correlation between the arms, so it biases the
estimates built on the second moments. The drop of the correlation coefficient
and the instability of the arm ratio $t$, which even inverts which arm is the
smaller against the stable $t\approx0.94$ of the other acquisitions, are two
manifestations of the same fact. This gives $\eta=\SI{9.0(3)}{\percent}$ (Fig.~\ref{fig:twb_R}, left), close
to the $\SI{8.5(7)}{\percent}$ reported in Ref.~\cite{razzoli:hybrid-2025}. This
global efficiency includes the collection and fiber-coupling losses of the
optical path, which is why it lies well below the nominal efficiency of the SiPM
alone (peak \SI{30}{\percent} at \SI{660}{\nano\meter},
Sec.~\ref{sec:setup-detection}). The
remaining four points all sit on the flat part of the curve, so they do not
constrain its rise toward $R=1$ at low $\langle m\rangle$, nor the background that
drives it.

As a second test, the lowest-intensity point is kept and the
$\langle m\rangle\approx2$ point, which sits slightly above the flat trend, is
excluded instead
(Fig.~\ref{fig:twb_R}, right). Although this is the noisier point, it is the only
measurement on the rising part of the curve, so keeping it lets the fit follow the
rise of $R$ rather than extrapolate it, and the resulting curve describes the
trend of the points better. This fit gives $\eta=\SI{9.1(2)}{\percent}$, in agreement with
the previous estimate, so the efficiency is robust against the choice of the
excluded point.

A third variant discards no point at all. It gives $\eta=\SI{8.8(2)}{\percent}$
with $\langle m_N\rangle=\num{0.092(10)}$ detected photons per shot, consistent
with the $\num{0.101(11)}$ of the second fit: whenever the lowest-intensity point
is kept the background is determined to about ten percent, while the first fit,
which drops that point, leaves it unconstrained at $\num{0.23(13)}$. The efficiency
moves by less than one standard deviation across the three fits, so it does not
depend on this choice, but the fitted background amounts to about \num{0.05}
counts per shot on each arm if shared equally, several times the dark-count level of
Sec.~\ref{sec:setup-detection}, and its origin cannot be identified with these
data. The value quoted above, which does not rely on that acquisition, is
therefore kept as the reference.

\section{The four-state alphabet}
\label{sec:alphabet}

The second measurement campaign implements the four-state protocol. A thermal
noise signal is added to a single arm of the twin beam, the signal arm, while the
other arm is left as the bare twin beam. Two bits per transmission are then encoded
in four states, labeled 00, 01, 10 and 11. The campaign
consists of nine acquisitions of $2\times10^{5}$ shots each, listed in
Table~\ref{tab:datasets}: the twin beam alone (dataset 0), the four noise signals
measured with the twin beam blocked (datasets 1, 3, 5, 7), and the four states,
twin beam plus noise, that carry the information (datasets 2, 4, 6, 8). The symbol
attached to each state reflects its position in the $(\langle m\rangle, R)$ plane
(Sec.~\ref{sec:constellation}), not the acquisition order, so in dataset order the
four states appear as 00, 01, 11, 10.

\begin{table}[htp]
	\centering
	\sisetup{separate-uncertainty=false}
	\caption{The nine acquisitions of the second campaign, $2\times10^{5}$ shots
	each. $\langle m_1\rangle$ and $\langle m_2\rangle$ are the mean detected
	photons in the two arms, $R$ the noise reduction factor and $\Gamma$ the
	correlation coefficient. In datasets 1, 3, 5, 7 the twin beam is blocked and
	only the noise is present on the signal arm ($\langle m_1\rangle\approx0$), so
	there $R>1$ (the Fano factor of the classical noise) and $\Gamma\approx0$ (the
	two arms uncorrelated); datasets 2, 4, 6, 8 carry twin beam plus noise and
	define the four states. Uncertainties are in compact form.}
	\label{tab:datasets}
	\begin{tabular}{S[table-format=1.0] S[table-format=1.2] S[table-format=1.2] S[table-format=1.3(1)] S[table-format=1.3(1)] l}
		\toprule
		{dataset} & {$\langle m_1\rangle$} & {$\langle m_2\rangle$} & {$R$} & {$\Gamma$} & {content} \\
		\midrule
		0 & 4.62 & 4.12 & 0.919(3) & 0.133(3) & TWB only \\
		1 & 0.01 & 0.51 & 1.196(6) & 0.000(2) & noise (00) \\
		2 & 4.62 & 4.59 & 0.930(3) & 0.126(2) & TWB + noise (00) \\
		3 & 0.01 & 0.54 & 1.414(5) & 0.003(3) & noise (01) \\
		4 & 4.63 & 4.60 & 0.946(3) & 0.122(3) & TWB + noise (01) \\
		5 & 0.00 & 0.66 & 1.538(7) & 0.001(3) & noise (11) \\
		6 & 4.64 & 4.70 & 0.951(2) & 0.124(2) & TWB + noise (11) \\
		7 & 0.00 & 0.68 & 1.246(5) & 0.003(2) & noise (10) \\
		8 & 4.62 & 4.70 & 0.939(3) & 0.123(2) & TWB + noise (10) \\
		\bottomrule
	\end{tabular}
\end{table}

\subsection{The four noise signals}
\label{sec:noise-signals}

\begin{figure}[htp]
	\centering
	\includegraphics[width=.8\textwidth]{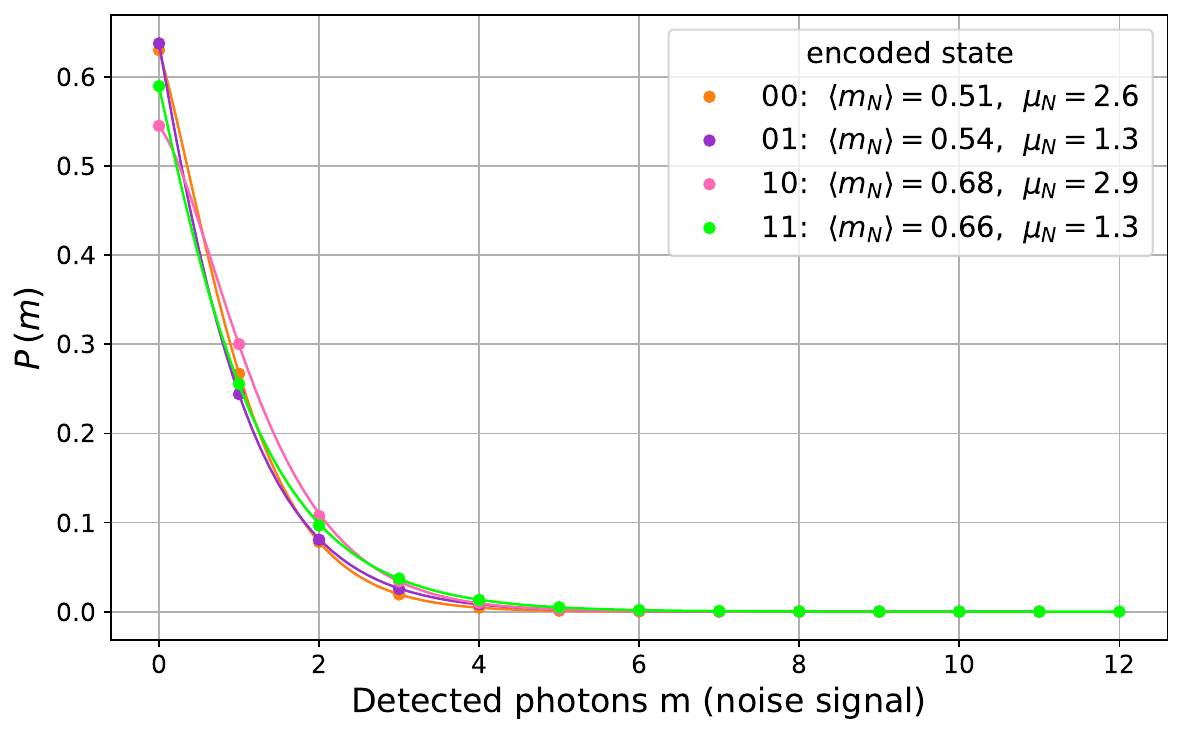}
	\caption{Detected-photon distributions of the four noise signals (markers) with
	the multi-mode thermal fits (lines), colored by the encoded state.}
	\label{fig:noise}
\end{figure}

The four noise signals are built by combining two mean values and two mode
numbers. Each is fitted with the multi-mode thermal distribution of
Eq.~\eqref{eq:mmtd} to extract the mean $\langle m_N\rangle$ and the number of
modes $\mu_N$, shown in Fig.~\ref{fig:noise}. The two mean values
($\langle m_N\rangle\approx0.52$ and
\num{0.67}) set the first bit through the mean value of the state, while the two mode
numbers ($\mu_N\approx1.3$ and \num{2.8}) set the second bit through the noise
reduction factor, since a smaller number of modes gives a larger $R$
(Eq.~\eqref{eq:R-model}). All signals are weak, with less than one detected photon
on average. With the twin beam blocked, the noise reduction factor of each signal
reduces to its Fano factor, $R=1+\langle m_N\rangle/\mu_N>1$
(Table~\ref{tab:datasets}): the noise alone is classical (super-shot-noise) and the
two arms are uncorrelated ($\Gamma\approx0$), so the sub-shot-noise values found
for the four states in Sec.~\ref{sec:constellation} originate from the twin beam.

\subsection{The four-state constellation}
\label{sec:constellation}

Each state is a twin beam with one of the noise signals added, and is described by
the mean detected photons $\langle m\rangle$ on the signal arm and the noise
reduction factor $R$. The noise reduction factor stays below one for all four
states, with $R$ between \num{0.930} and \num{0.951}, so the nonclassical correlation
survives the added noise; the correlation coefficient is $\Gamma\approx0.12$,
stable across the alphabet.

\begin{figure}[htp]
	\centering
	\subfloat[]{\includegraphics[width=.8\textwidth]{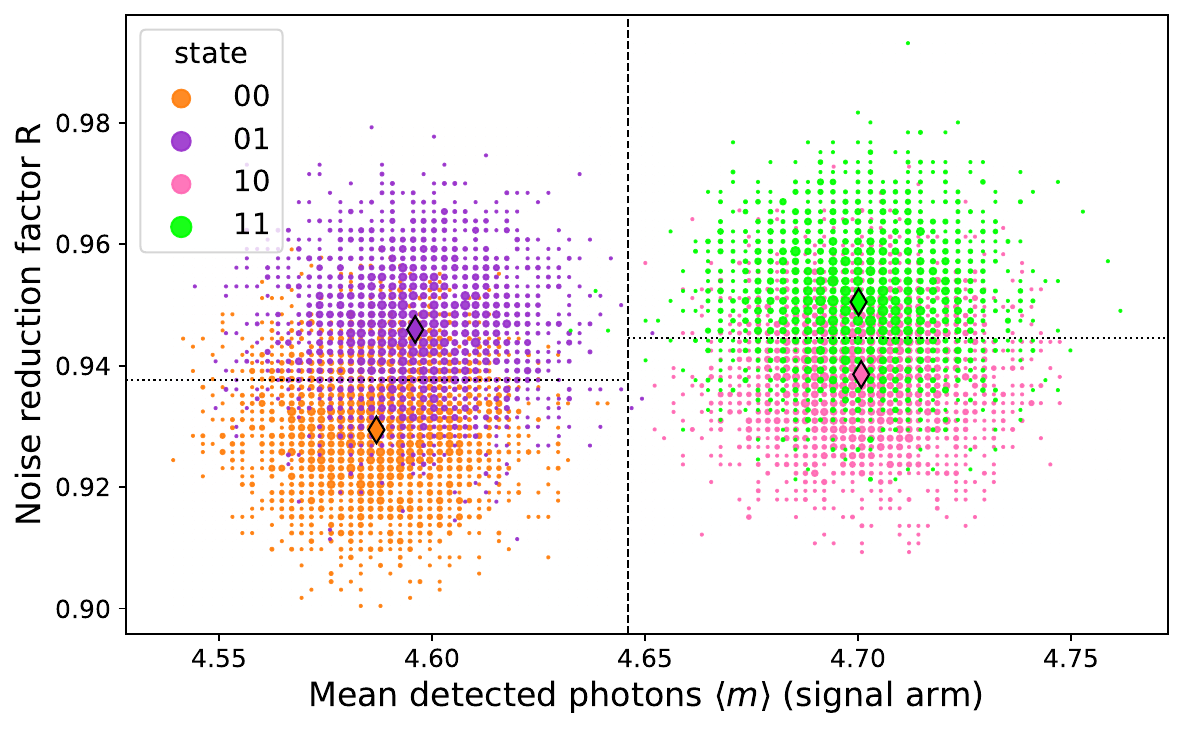}}\\
	\subfloat[]{\includegraphics[width=.8\textwidth]{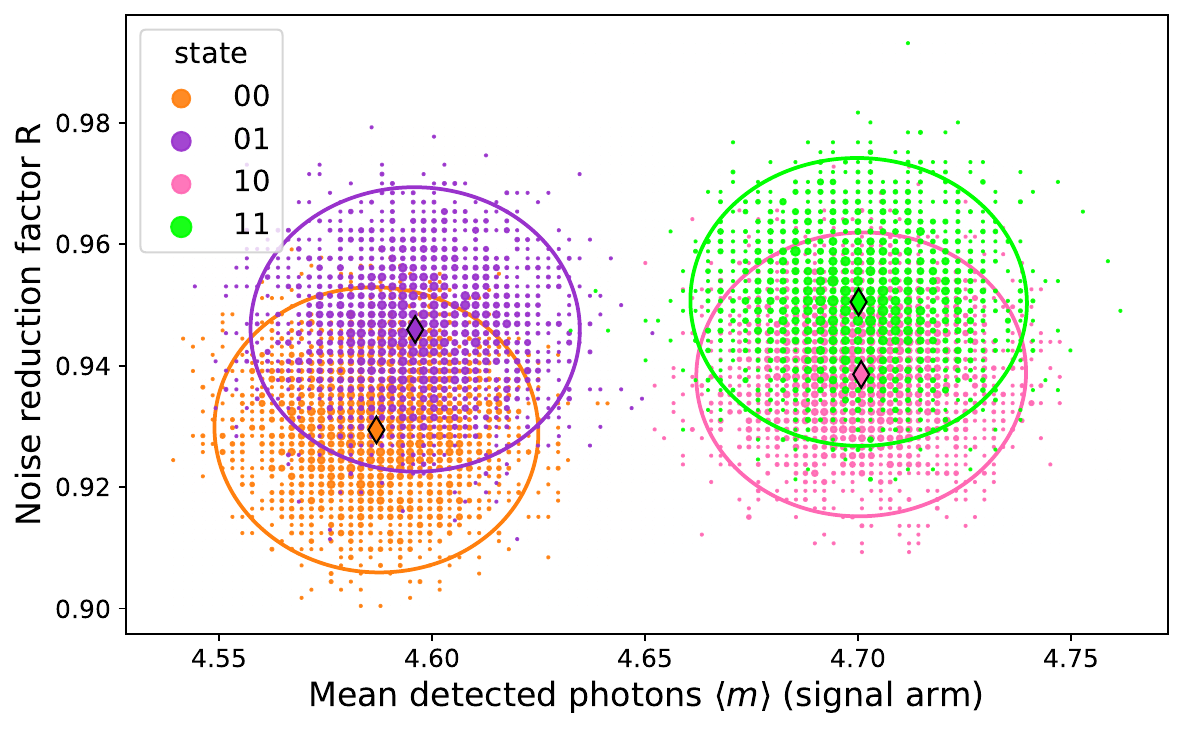}}
	\caption{The four states in the $(\langle m\rangle, R)$ plane, from bootstrap
	batches of \num{20000} shots. (a) The clouds with the order-zero decision
	thresholds (dashed: mean value; dotted: $R$ within each half). (b) The same
	clouds with the $95\%$ confidence ellipses. The state centroids are at
	$(\langle m\rangle, R) = (4.59, 0.930)$, $(4.60, 0.946)$, $(4.70, 0.939)$ and
	$(4.70, 0.951)$ for 00, 01, 10 and 11, respectively. The mean value separates
	the two pairs \{00, 01\} and \{10, 11\} cleanly, while $R$ separates the two
	states within each pair only partially, their clouds overlapping.}
	\label{fig:constellation}
\end{figure}

The four states occupy four points of the $(\langle m\rangle, R)$ plane. A reliable
estimate of $R$ requires a large sample, so to visualize the spread of a single
measurement each acquisition is resampled by bootstrap: \num{3000} batches of
\num{20000} shots are drawn with replacement, and each batch gives one point
$(\langle m\rangle, R)$. The resulting clouds of \num{3000} points are shown in
Fig.~\ref{fig:constellation}. Each cloud is rendered as a two-dimensional
histogram, with the marker area proportional to the number of estimates in each
bin. For each state the $95\%$ confidence region is drawn as an ellipse, built from the
$2\times2$ covariance matrix of the cloud: its eigenvectors give the directions of
the ellipse axes and its eigenvalues the variances along them. The scale of the
ellipse follows from the statistics of a two-dimensional Gaussian. The squared
distance of a point from the centroid, normalized by the covariance (the
Mahalanobis distance), follows a chi-squared distribution with two degrees of
freedom, one for each coordinate ($\langle m\rangle$ and $R$). The ellipse that
contains $95\%$ of the points is the one whose squared Mahalanobis radius equals
the $95\%$ quantile of this distribution, denoted $k$ ($k\approx5.99$); its
semi-axes are therefore the standard deviations along the principal directions
multiplied by $\sqrt{k}\approx2.45$. The Gaussian approximation is justified by the
central limit theorem, since each point is an average over a large batch.

The constellation shows the working principle of the discrimination. Along the
horizontal axis the two pairs are well separated, whereas along $R$ the two states
of each pair overlap. This suggests an order-zero decoding
(Fig.~\ref{fig:constellation}a): a threshold on the mean value assigns the first
bit, and a threshold on $R$ within each half assigns the second. This threshold
decoding is the basis of the protocol described in Sec.~\ref{sec:protocol}.

\subsection{Comparison with the model}
\label{sec:R-model-compare}

\begin{figure}[htp]
	\centering
	\includegraphics[width=.8\textwidth]{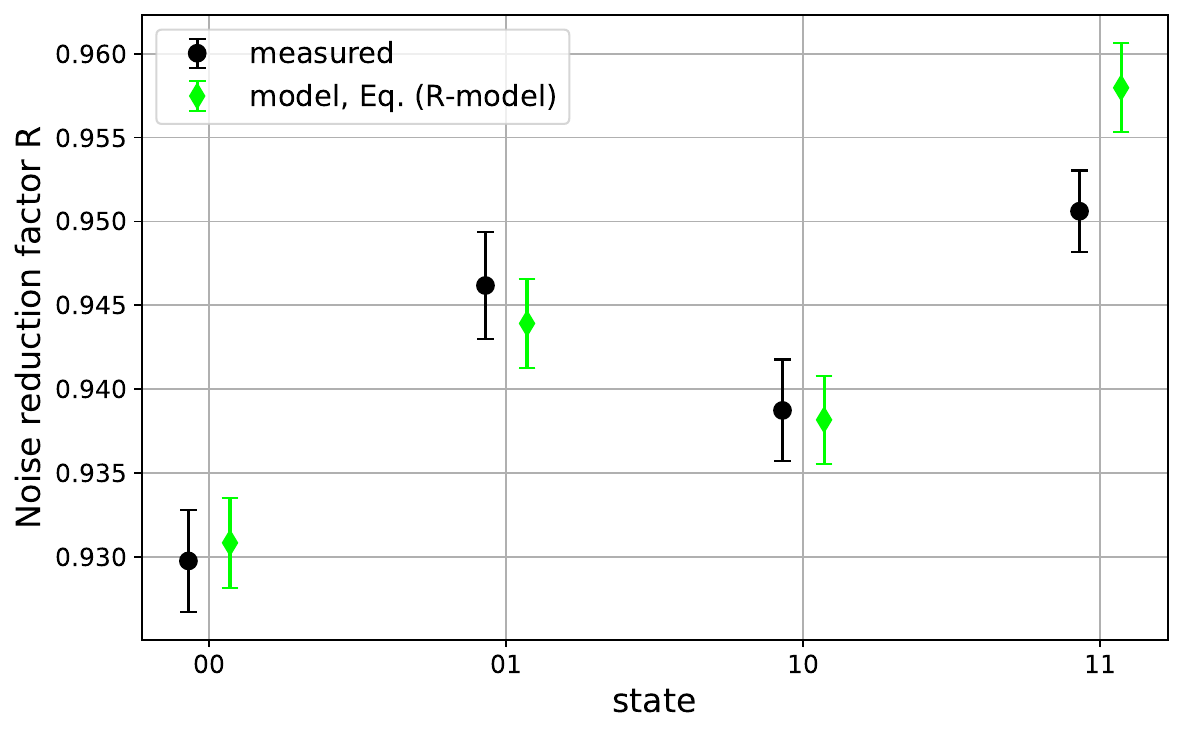}
	\caption{Measured noise reduction factor of the four states (circles) and the
	prediction of Eq.~\eqref{eq:R-model} (diamonds), with the error bar on the
	prediction propagated from $\eta$, $t$, $\mu$ and $\mu_N$. Three states agree with
	the prediction within one standard deviation, while state 11 shows a mild tension.}
	\label{fig:Rmodel}
\end{figure}

The measured noise reduction factor of the four states is compared with the
prediction of Eq.~\eqref{eq:R-model} in Fig.~\ref{fig:Rmodel}. The model is
evaluated with the detection efficiency $\eta=\SI{9.0(3)}{\percent}$ from Part~I,
the imbalance $t=\num{0.890(2)}$ measured on dataset~0 (the twin beam alone of this
campaign), the measured mean values, and the mode numbers of the twin beam and of
the noise, $\mu=\num{71(4)}$ and $\mu_N$ (Fig.~\ref{fig:noise}), from the multi-mode
thermal fits. This $t$ is recomputed from the twin-beam reference of the present
campaign and differs slightly from the value $t\approx0.94$ of Part~I, which is
measured on a separate acquisition.

The uncertainty on the prediction is obtained by Monte Carlo propagation. Each
input ($\eta$, $t$, $\mu$, $\mu_N$) is sampled $\num{40000}$ times from a Gaussian
centered on its measured value with width equal to its uncertainty, the model is
evaluated for every draw, and the standard deviation of the resulting distribution
of $R$ is taken as the error bar on the prediction. This replaces the analytic
propagation of the uncertainties, which is impractical for a nonlinear expression
such as Eq.~\eqref{eq:R-model}, and requires only that the inputs and their errors
be known.

Three of the four states agree with the prediction within one standard deviation,
while for state 11 the model overestimates $R$ by about two standard deviations.
This state combines a small number of modes ($\mu_N\approx1.3$) with the larger
of the two noise means, so its noise term $\langle m_N\rangle^{2}/\mu_N$ in
Eq.~\eqref{eq:R-model} is the largest of the four states and the prediction is
the most sensitive to $\mu_N$. The same offset appears in
Sec.~\ref{sec:additivity} without any model, where it is traced to the twin-beam
reference: the twin beam carried by this state is better correlated than the one
of dataset~0, from which the model inputs are taken. The residual discrepancy is compatible
with the simplified description of the channel by a single efficiency and a single
imbalance factor.

\subsection{The states as twin beam plus independent noise}
\label{sec:additivity}

The additive picture behind the model can also be checked directly on the data,
without any fit. The pure twin beam (dataset~0) and each noise signal are added
shot by shot on the signal arm, and the single-arm Fano factor $F$ and the noise
reduction factor $R$ of the resulting synthetic state are compared with the
measured ones (Table~\ref{tab:additivity}). The noise is a thermal signal from a
separate optical path, statistically independent of the twin beam, so on the signal
arm the means and the variances of the two contributions add: the signal arm is
mildly super-Poissonian ($F$ between \num{1.07} and \num{1.11}) while the pair stays
sub-shot-noise.

The Fano factor is reproduced almost exactly by the addition, which confirms that
the marginal statistics of the signal arm are those of the twin beam plus the
independent noise. The correlation $R$ is reproduced within about one percent,
slightly overestimated and by the largest amount for state 11, the same behavior
found for the model in Sec.~\ref{sec:R-model-compare}: the residual is therefore
not an artifact of the model assumptions but is already present in the raw counts.
The residual comes from the reference itself: dataset~0 is a separate acquisition,
slightly brighter and slightly less correlated ($R=\num{0.919}$) than the twin beam
carried by the states, which the same additive decomposition puts between
\num{0.906} and \num{0.917}. The lowest value belongs to state 11, which is why its
residual is the largest. Replacing only that value reproduces the added column,
so summing the two contributions after the fact does not degrade the state by
itself. Within this limitation the four states are consistent with a twin beam plus
an independent thermal noise.

\begin{table}[htp]
	\centering
	\sisetup{separate-uncertainty=false}
	\caption{Single-arm Fano factor $F$ of the signal arm and noise reduction factor
	$R$ of the four states, measured and obtained by adding the pure twin beam
	(dataset~0) and the noise signal shot by shot on the signal arm. Uncertainties
	are in compact form.}
	\label{tab:additivity}
	\begin{tabular}{c S[table-format=1.4(2)] S[table-format=1.4(2)] S[table-format=1.4(2)] S[table-format=1.4(2)]}
		\toprule
		{state} & {$F$ (measured)} & {$F$ (added)} & {$R$ (measured)} & {$R$ (added)} \\
		\midrule
		00 & 1.0718(21) & 1.0760(21) & 0.9298(30) & 0.9353(28) \\
		01 & 1.1010(35) & 1.1019(25) & 0.9462(32) & 0.9497(33) \\
		10 & 1.0797(35) & 1.0874(20) & 0.9387(30) & 0.9428(35) \\
		11 & 1.1128(28) & 1.1261(29) & 0.9506(24) & 0.9606(39) \\
		\bottomrule
	\end{tabular}
\end{table}

\section{Encoding and decoding}
\label{sec:protocol}

The constellation of Sec.~\ref{sec:constellation} defines a four-symbol alphabet:
each state occupies a distinct point of the $(\langle m\rangle, R)$ plane. This
alphabet is here used to build a communication protocol. In the communication picture Alice prepares the
twin beam and encodes each symbol on the signal arm she sends to the receiver Bob,
keeping the idler arm as a reference. Two bits are carried per
transmission, one in the mean value and one in the noise reduction factor, and each
symbol is sent as a block of $N$ shots. The receiver reads back the two bits from
the $(\langle m\rangle, R)$ of the block through a set of fixed thresholds. The
noise reduction factor also provides the security check: an eavesdropper who
intercepts part of the signal and resends uncorrelated light leaves the mean value
unchanged but raises $R$, so the attack becomes visible, as discussed later.

\subsection{Encoding}
\label{sec:encoding}

The message is first written as a bit string, using the eight-bit ASCII code of
each character, and the bits are grouped two at a time into symbols. The
four symbols 00, 01, 10 and 11 are the four states of the alphabet, so the scheme is
a quaternary encoding that carries two bits per symbol. Each symbol is transmitted
as a block of $N$ shots of the corresponding state. In a real link each block would
be a fresh acquisition of $N$ shots; here, with a single finite acquisition
available per state, a block is instead emulated by bootstrap, drawing $N$ shots
with replacement from that acquisition, using the same random indices on the two
arms so that the twin-beam correlation between signal and idler is preserved.
With replacement, a block of $N$ draws from an acquisition of the same size does not
reproduce it exactly. A given shot is absent from it with probability
\begin{equation}
	\label{eq:bootstrap-fraction}
	\left(1-\frac{1}{N}\right)^{N}\xrightarrow[N\to\infty]{}\frac{1}{e}
\end{equation}
a standard limit, so at $N=\num{2e5}$ about $1/e\approx37\%$ of the shots are missing
and $1-1/e\approx63\%$ appear, some of them more than once. Each block is therefore a different reweighting of the same data,
and it is this block-to-block variability that spreads the points around the state
centroids in Fig.~\ref{fig:protocol} rather than collapsing them onto a single
point. In this way the bootstrap emulates a stream of transmissions from a single
finite acquisition, although not fully independent ones: of the shots of one block,
the fraction that also appear in another follows from the same limit, and it is
$1-1/e\approx\num{63}\%$ when each block is as long as the acquisition, falling to
about the ratio of the two lengths for much shorter blocks. Independent
transmissions therefore require $N$ of the order of a few thousand out of the
\num{2e5} shots available per state, for an overlap of one percent, far fewer than
the blocks used below.
Resampling a fixed dataset remains a simplification,
and its effect is examined with the worked example (Sec.~\ref{sec:message}). The block size $N$ is the
resource spent per symbol: the larger it is, the better $\langle m\rangle$ and $R$
are estimated and the more reliable the decoding, but the lower the transmission
rate, since more shots are used for each symbol. Small blocks also weaken the
security check, because the test that reveals an eavesdropper is itself an estimate
built on $R$ and its power grows with the block size
(Sec.~\ref{sec:attack-detection}). The trade-off is therefore between rate on one
side and both reliability and security on the other, and it is quantified in the
next section.

\subsection{Decoding}
\label{sec:decoding}

The receiver splits the stream into blocks of $N$ shots and, for each block,
estimates the mean detected photons on the signal arm and the noise reduction
factor. The mean value uses only the signal arm, but the noise reduction factor is
a joint quantity of the two arms (Eq.~\eqref{eq:R}): Bob therefore also needs
Alice's idler counts, sent to him shot by shot over a classical channel. Without
them only the mean value, and hence the first bit, can be recovered. The decoding is order zero, by which we mean the simplest possible rule:
two fixed thresholds applied in sequence (Table~\ref{tab:decode}). Each bit is set
to 0 when its quantity is below the threshold and to 1 when above. The first (most
significant) bit comes from the mean value, with $m_{\mathrm{th}}\approx\num{4.65}$
separating the two pairs \{00, 01\} and \{10, 11\}. The second bit comes from $R$,
read the same way in both pairs, with $R_{\mathrm{th}}^{L}\approx\num{0.938}$ on the
left and $R_{\mathrm{th}}^{R}\approx\num{0.945}$ on the right. All thresholds are the
midpoints between the state centroids, so the decision map is the one already shown
in Fig.~\ref{fig:constellation}a.

The two states of each pair, which are
the ones that overlap and can be confused, differ by a single bit, so a misread
symbol costs one bit and the bit error rate is about half the symbol error rate. The
first bit, set by the well-separated mean values, is essentially always correct, and
almost all errors fall on the second bit, the one read from $R$. The receiver is
assumed to know the block size and the block boundaries, that is the stream is
synchronized.

\subsection{A transmitted message}
\label{sec:message}

\begin{figure}[htp]
	\centering
	\subfloat[]{\includegraphics[width=0.8\textwidth]{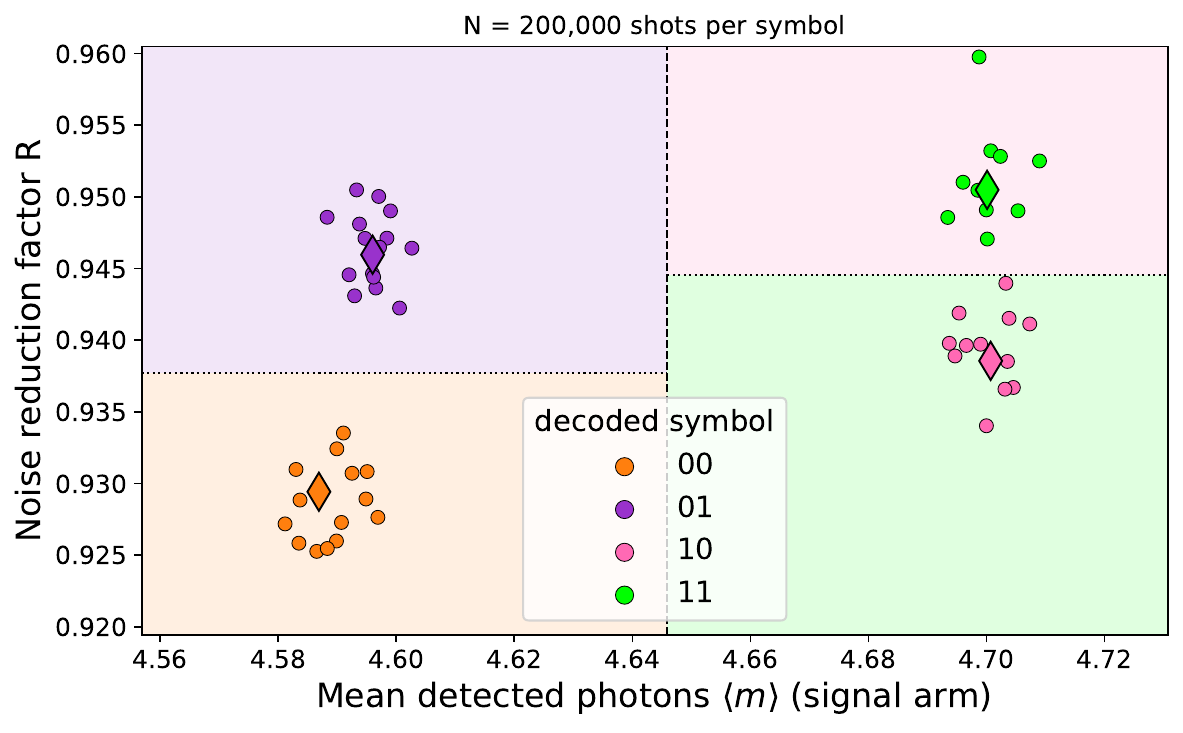}} \\
	\subfloat[]{\includegraphics[width=0.8\textwidth]{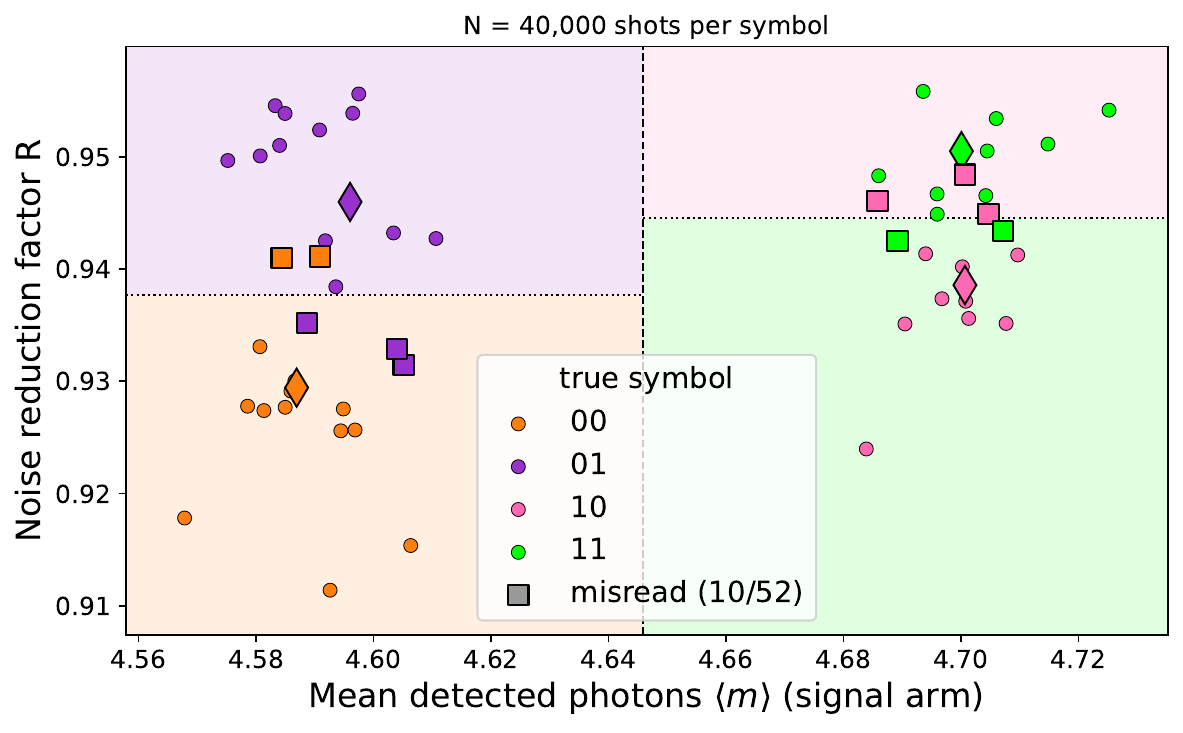}}
	\caption{The message ``Hello, world!'' (52 symbols) transmitted and decoded.
	Each point is one block, placed at its estimated $(\langle m\rangle, R)$, over
	the four decision regions; diamonds mark the state centroids. (a) With
	$N=\num{2e5}$ shots per symbol the blocks sit well inside their regions, colored
	by the decoded symbol, and the message is recovered exactly. (b) With
	$N=\num{4e4}$ the blocks scatter across the thresholds; here they are colored by
	the true symbol, and the misread ones, which fall in a region of a different
	color, are drawn as squares.}
	\label{fig:protocol}
\end{figure}

Figure~\ref{fig:protocol} shows the protocol on the message ``Hello, world!'',
which is 52 symbols long. The block size is first set deliberately large,
$N=\num{2e5}$ shots per symbol, so that the estimates are precise enough to leave no
errors: every block falls well inside the decision region of its state and the
message is recovered exactly (Fig.~\ref{fig:protocol}a). It is then lowered to a
more realistic $N=\num{4e4}$, where the estimates of $\langle m\rangle$ and $R$
scatter much more, several blocks cross the thresholds into a neighboring region,
and the decoded text comes out corrupted (Fig.~\ref{fig:protocol}b).

The error-free recovery at large $N$ should be read with care. The transmitted
blocks are resampled from the same acquisition used to fix the thresholds, so as
$N$ grows the block estimates converge to the per-state values of that acquisition,
which lie on the correct side of the thresholds by construction. In a real link the
thresholds would be fixed once on a calibration acquisition and the protocol run on
different data, so the large-$N$ limit would not be error-free. The meaningful
quantity is the error rate at finite $N$, studied systematically as a function of
the block size in the next section.

\begin{table}[htp]
	\centering
	\caption{Order-zero decision rule. $m_{\mathrm{th}}\approx\num{4.65}$ is the
	threshold on the mean value; $R_{\mathrm{th}}^{L}\approx\num{0.938}$ and
	$R_{\mathrm{th}}^{R}\approx\num{0.945}$ are the thresholds on the noise reduction
	factor in the left and right pair. The $R$ test has the same direction in both
	pairs.}
	\label{tab:decode}
	\begin{tabular}{c c c}
		\toprule
		symbol & mean value & noise reduction factor \\
		\midrule
		00 & $\langle m\rangle < m_{\mathrm{th}}$ & $R < R_{\mathrm{th}}^{L}$ \\
		01 & $\langle m\rangle < m_{\mathrm{th}}$ & $R \ge R_{\mathrm{th}}^{L}$ \\
		10 & $\langle m\rangle \ge m_{\mathrm{th}}$ & $R < R_{\mathrm{th}}^{R}$ \\
		11 & $\langle m\rangle \ge m_{\mathrm{th}}$ & $R \ge R_{\mathrm{th}}^{R}$ \\
		\bottomrule
	\end{tabular}
\end{table}

\section{Error analysis}
\label{sec:errors}

The worked example of Sec.~\ref{sec:message} shows single realizations of the
protocol. This section measures its error rate systematically as a function of
the block size $N$, the resource spent per symbol, extending to the four-state
alphabet the batch-size analysis carried out for the binary protocol in
Ref.~\cite{razzoli:hybrid-2025}. One methodological point is added: the data used
to calibrate the decoder are kept disjoint from the data used to test it, so the
results are free from the self-consistency artifact discussed in
Sec.~\ref{sec:message}.

\subsection{Disjoint calibration and test}
\label{sec:err-method}

Each \num{2e5}-shot acquisition is split into two halves of \num{e5} shots:
the even-index shots form the calibration set and the odd-index shots the test
set. The interleaved split makes the two halves insensitive to slow drifts of the
source; contiguous halves were also checked and give compatible statistics. The
order-zero thresholds of Sec.~\ref{sec:decoding} are recomputed on the
calibration set alone, giving $m_{\mathrm{th}}=\num{4.645}$,
$R_{\mathrm{th}}^{L}=\num{0.935}$ and $R_{\mathrm{th}}^{R}=\num{0.945}$, close to
the values obtained from the full dataset. Blocks of $N$ shots are then emulated
by bootstrap as in Sec.~\ref{sec:encoding}, drawing with replacement from the
test set only, with the same indices on the two arms, and each block is decoded
with the calibrated thresholds; \num{3000} blocks per state are generated for each
$N$ between \num{e3} and \num{3.2e5}. Above the pool size of \num{e5} shots the
resampled blocks reuse each shot several times: drawing with replacement still
emulates longer blocks with the same empirical distribution, under the assumption
of independent shots. The four symbols are taken as
equiprobable, the standard neutral choice when characterizing a channel, so the
symbol error rate (SER), the fraction of blocks decoded as a wrong symbol, is the
unweighted average of the per-state error rates; the bit error rates
(BER) count instead the wrong bits, and BER$_1$, BER$_2$ refer separately to the
first (mean value) and second ($R$) bit. A specific message would instead weight
the per-state rates, which differ considerably (Sec.~\ref{sec:err-confusion}), by
its own symbol frequencies. The quoted uncertainties are binomial standard errors
over the emulated blocks, so they reflect the finite number of resamplings only.
The dominant uncertainty is the fluctuation of the single acquisition from which
all blocks are drawn: exchanging the roles of the calibration and test halves
moves the symbol error rate at $N=\num{4e4}$ from \num{0.235} to \num{0.162}, a
shift about twenty times larger than the binomial errors. The same fluctuation
governs the slow decay of the error rate at large $N$ discussed below.

\subsection{Error probability versus block size}
\label{sec:err-vs-N}

\begin{figure}[htp]
	\centering
	\includegraphics[width=.8\textwidth]{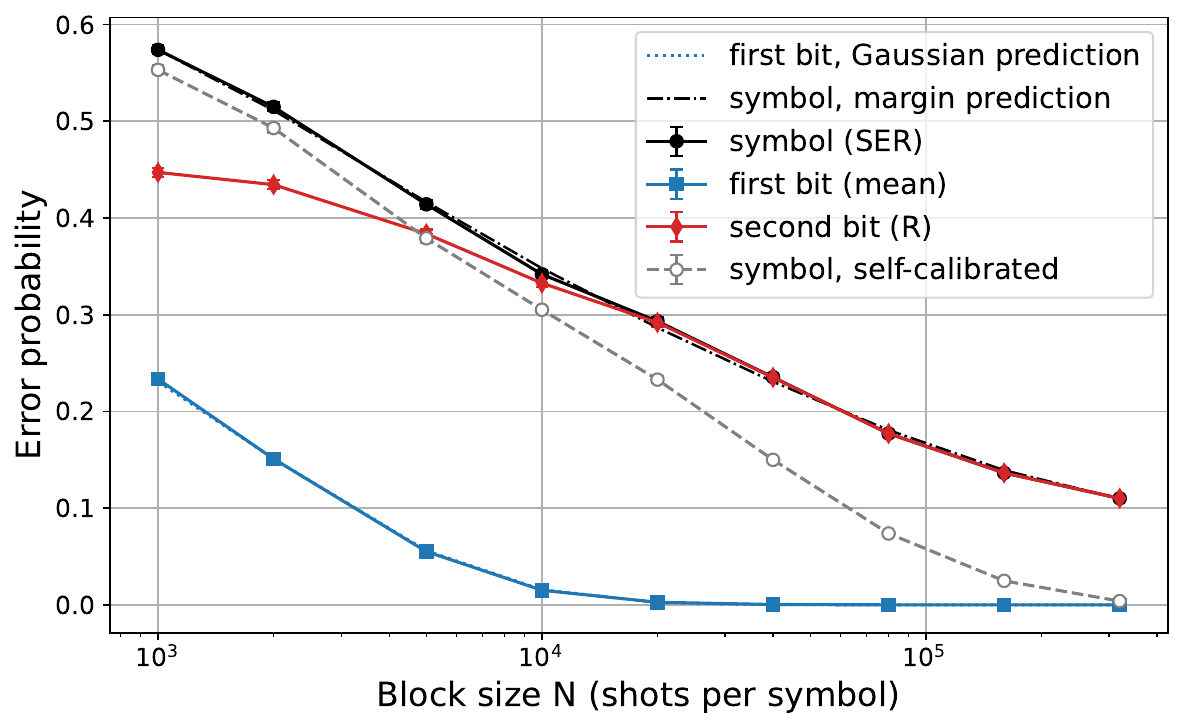}
	\caption{Error probability as a function of the block size $N$, with disjoint
	calibration and test data: symbol error rate and error rates of the two bits.
	The dotted line is the Gaussian prediction for the first bit and the
	dash-dotted line is the margin prediction for the symbol error rate, both
	described in Sec.~\ref{sec:err-vs-N}. The dashed
	line with open markers is the self-calibrated symbol error rate, with
	thresholds and blocks from the same dataset as in Sec.~\ref{sec:message},
	which decays to zero by construction. Error bars are binomial standard errors.}
	\label{fig:errN}
\end{figure}

Figure~\ref{fig:errN} and Table~\ref{tab:errors} collect the results. The symbol
error rate decreases from \num{0.57} at $N=\num{e3}$, where the second bit is
barely better than a coin flip ($\mathrm{BER}_2=\num{0.447}$), to \num{0.235} at
$N=\num{4e4}$ and \num{0.110} at
$N=\num{3.2e5}$. The two bits behave very
differently. The first bit follows plain Gaussian statistics. For blocks of a given state the
estimated mean is Gaussian, centered on the state mean $\langle m\rangle$ with
standard deviation $\sigma/\sqrt{N}$, where $\sigma^{2}$ is the single-shot
variance of the detected photons; the bit is wrong when the estimate falls on the
wrong side of the calibrated threshold $m_{\mathrm{th}}$, which happens with
probability $\Phi(-|\langle m\rangle-m_{\mathrm{th}}|\sqrt{N}/\sigma)$, with
$\Phi$ the cumulative distribution function of the standard Gaussian and
$|\langle m\rangle-m_{\mathrm{th}}|$ the margin of the state from the threshold.
This prediction, averaged over the four states and shown as the dotted curve in
Fig.~\ref{fig:errN}, reproduces the measured BER$_1$ at every $N$, and the first
bit becomes essentially error-free for $N\gtrsim\num{2e4}$. The second bit dominates the error budget at all
block sizes; since the two states of each pair differ by that single bit, the bit
error rate is about half the symbol error rate, as anticipated in
Sec.~\ref{sec:decoding}.

\begin{figure}[htp]
	\centering
	\includegraphics[width=.8\textwidth]{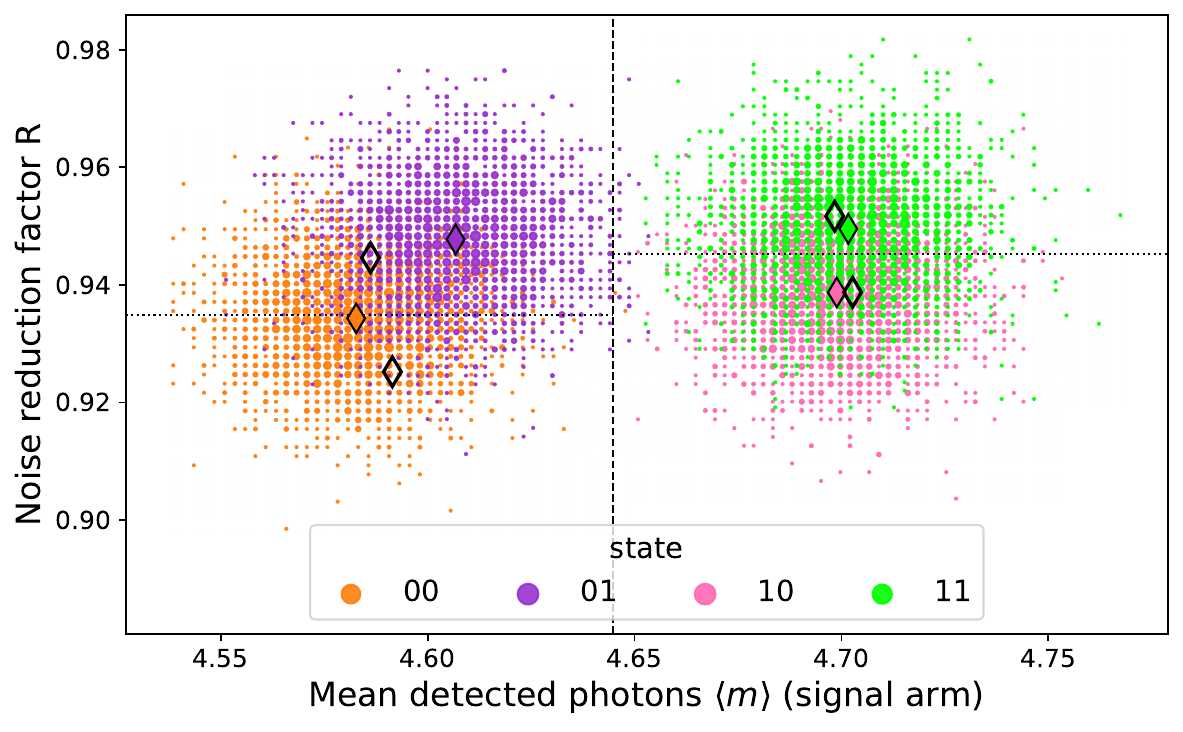}
	\caption{Bootstrap clouds of the test pools at $N=\num{2e4}$, the batch size of
	the constellation of Fig.~\ref{fig:constellation}, with the thresholds
	calibrated on the other half of the data. Filled diamonds: test centroids;
	open diamonds: calibration centroids. The calibrated $R$ threshold cuts the 00
	cloud through the middle, its centroid within \num{0.0006} of the line, while
	its partner 01 sits clear of it: this mismatch drives the slow decay of
	Fig.~\ref{fig:errN}. In Fig.~\ref{fig:constellation}a, by contrast, thresholds
	and clouds come from the same data.}
	\label{fig:testclouds}
\end{figure}

The dashed curve with open markers shows, for comparison, the self-calibrated
configuration of Sec.~\ref{sec:message}, with thresholds and blocks drawn from
the same full dataset: it keeps decaying (\num{0.150} at $N=\num{4e4}$ and
\num{0.004} at $N=\num{3.2e5}$) because the block statistics converge to the very centroids
that define the thresholds. With disjoint data the decay is much slower, and its
pace is set by the mismatch between the calibrated thresholds and the test data.
The noise reduction factor is a variance
divided by a mean (Eq.~\eqref{eq:R}), and the relative sampling error of a
variance estimated on $N$ nearly Gaussian samples is $\sqrt{2/N}$ (the sample
variance is the average of $N$ squared deviations, each of mean $\sigma^{2}$
and, for Gaussian data, variance $2\sigma^{4}$), so the spread
of its estimates is expected to scale as
$\sigma(R)\simeq R\sqrt{2/N}\approx1.3/\sqrt{N}$. The emulated blocks confirm
this: the measured $\sigma(R)\sqrt{N}\approx\num{1.35}$ is constant over the
whole grid for all four states. The $R$ centroids of the calibration and test
sets, each estimated on \num{e5} shots, therefore differ by amounts of order
\num{0.004}, comparable to the decision margins in $R$ (the calibrated
thresholds sit \num{0.010} and \num{0.007} from the calibration centroids in the
left and right pair). All four test centroids happen to
lie on the correct side of the calibrated thresholds, so even with disjoint data
the error rate eventually vanishes, but the margins set the pace: for each state
the probability of a wrong symbol is $1-(1-p_1)(1-p_2)$, where
$p_k=\Phi(-\Delta_k\sqrt{N}/\sigma_k)$ is the probability of crossing threshold
$k$, $\Delta_k$ the distance of the test centroid from that threshold, and
$\sigma_k$ the corresponding single-shot spread ($\sigma$ for the mean,
\num{1.35} for $R$). This margin prediction, the dash-dotted curve of
Fig.~\ref{fig:errN}, is built only from the measured centroids, thresholds and
spreads, and reproduces the measured rate over the whole grid. In the present split the test centroid of state 00 falls at
$R=\num{0.9343}$, within \num{0.0006} of the calibrated threshold, while its
partner 01 sits a comfortable \num{0.0128} away on the other side: the second bit
of state 00 remains a near coin flip until $\sqrt{N}\gg 1.35/0.0006$, that is
$N\sim10^{7}$, far beyond the acquisition itself. Figure~\ref{fig:testclouds}
makes the mismatch visible: the calibrated $R$ threshold runs through the middle
of the 00 test cloud, while it only clips the lower tail of the 01 one. The margins, and with them the
achievable error rate, are a property of the calibration realization rather than
of the protocol,
as the exchange of the two halves quoted in Sec.~\ref{sec:err-method} already
shows; calibrations restricted to \num{2.5e4}, \num{5e4} and \num{e5} shots give
a symbol error rate at $N=\num{4e4}$ of
\num{0.19}, \num{0.24} and \num{0.23}, with no clear trend: the fluctuations of
the single realization dominate. A practical implementation would therefore need
calibration samples large enough to make the threshold uncertainty small against
the margins, or a periodic recalibration of the thresholds.

\begin{table}[htp]
	\centering
	\sisetup{separate-uncertainty=false}
	\caption{Error rates, AUC and mutual information as a function of the block
	size $N$. SER is the symbol error rate, BER$_1$ and BER$_2$ the error rates of
	the two bits, AUC$_\mathrm{L}$ and AUC$_\mathrm{R}$ quantify the $R$
	discrimination within the left and right pair (Sec.~\ref{sec:err-roc}), and
	$I(X;Y)$ is the mutual information between sent and decoded symbol in bits per
	symbol (Sec.~\ref{sec:err-info}). The uncertainties on SER, BER$_1$ and BER$_2$ are binomial standard errors
	over the \num{12000} emulated blocks, in compact form; AUC and $I(X;Y)$ are
		quoted without an error estimate.}
	\label{tab:errors}
	\begin{tabular}{S[table-format=6.0] S[table-format=1.4(2)] S[table-format=1.4(2)] S[table-format=1.4(2)] S[table-format=1.3] S[table-format=1.3] S[table-format=1.3]}
		\toprule
		{$N$} & {SER} & {BER$_1$} & {BER$_2$} & {AUC$_\mathrm{L}$} & {AUC$_\mathrm{R}$} & {$I(X;Y)$} \\
		\midrule
		1000 & 0.5738(45) & 0.2334(39) & 0.4471(45) & 0.592 & 0.566 & 0.231 \\
		2000 & 0.5148(46) & 0.1509(33) & 0.4344(45) & 0.617 & 0.588 & 0.415 \\
		5000 & 0.4141(45) & 0.0551(21) & 0.3836(44) & 0.688 & 0.654 & 0.749 \\
		10000 & 0.3417(43) & 0.0151(11) & 0.3327(43) & 0.762 & 0.727 & 0.990 \\
		20000 & 0.2932(42) & 0.0027(5) & 0.2918(41) & 0.846 & 0.778 & 1.124 \\
		40000 & 0.2353(39) & 0.0003(2) & 0.2351(39) & 0.918 & 0.869 & 1.249 \\
		80000 & 0.1771(35) & 0.0000(1) & 0.1771(35) & 0.979 & 0.945 & 1.395 \\
		160000 & 0.1364(31) & 0.0000(1) & 0.1364(31) & 0.997 & 0.991 & 1.531 \\
		320000 & 0.1101(29) & 0.0000(1) & 0.1101(29) & 1.000 & 0.999 & 1.632 \\
		\bottomrule
	\end{tabular}
\end{table}

\subsection{Confusion matrices}
\label{sec:err-confusion}

\begin{figure}[htp]
	\centering
	\includegraphics[width=\textwidth]{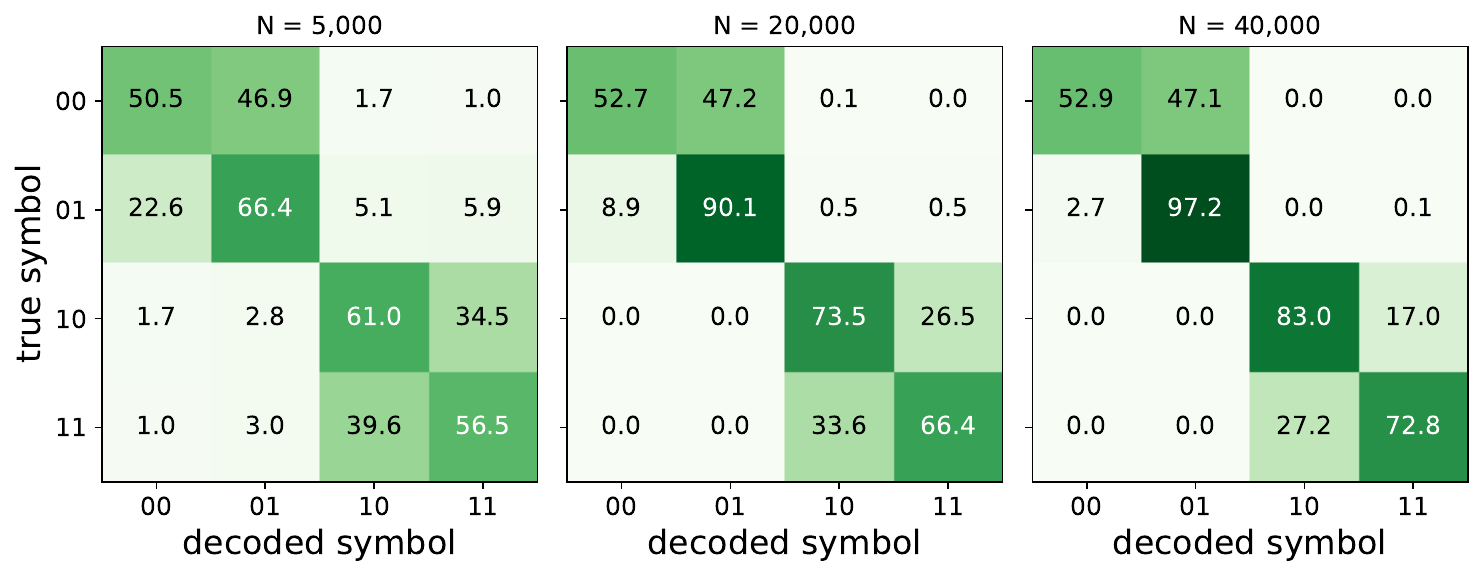}
	\caption{Confusion matrices (rows: sent symbol; columns: decoded symbol;
	entries in percent, row-normalized) at three block sizes. The residual
	confusion at large $N$ is entirely within the two pairs and concentrates on
	state 00, whose test centroid sits on the calibrated $R$ threshold.}
	\label{fig:confusion}
\end{figure}

Figure~\ref{fig:confusion} details how the errors distribute among the symbols at
three block sizes. At $N=\num{5e3}$ both bits fail: a few percent of the
blocks cross the mean threshold into the other pair, and the two states of each
pair mix heavily. At $N=\num{2e4}$ the two pairs no longer mix, so the first
bit is exact and all residual errors are within the pairs; every misread symbol
then costs a single bit, which keeps the bit error rate at half the symbol error
rate. At $N=\num{4e4}$ the intra-pair confusion decreases further for three
states, while state 00 remains decoded correctly only slightly above half of the
time, the signature of the calibration mismatch discussed in
Sec.~\ref{sec:err-vs-N}. The strong asymmetry within the left pair reflects the
very different margins of its two states from the calibrated threshold, visible
directly in Fig.~\ref{fig:testclouds}: the errors concentrate on state 00, and
with the swapped split of Sec.~\ref{sec:err-method} the roles reverse, with
state 01 misread instead.

\subsection{ROC curves of the second bit}
\label{sec:err-roc}

Within each pair the second bit is a binary discrimination based on $R$, and its
quality can be assessed independently of any specific threshold with a receiver
operating characteristic (ROC): sweeping the decision threshold over all values
traces the true-positive versus false-positive rates, taking the high-$R$ state
of the pair (01 or 11) as the positive class. The area under the curve (AUC) is
the probability that a block of the high-$R$ state yields a larger $R$ than a
block of the low-$R$ state, computed with the Mann-Whitney statistic; \num{0.5}
is chance level and \num{1} perfect discrimination. Being threshold-independent,
the AUC measures the achievable separation of the $R$ distributions, regardless
of the calibration mismatch that limits the fixed-threshold decoder.

\begin{figure}[htp]
	\centering
	\includegraphics[width=\textwidth]{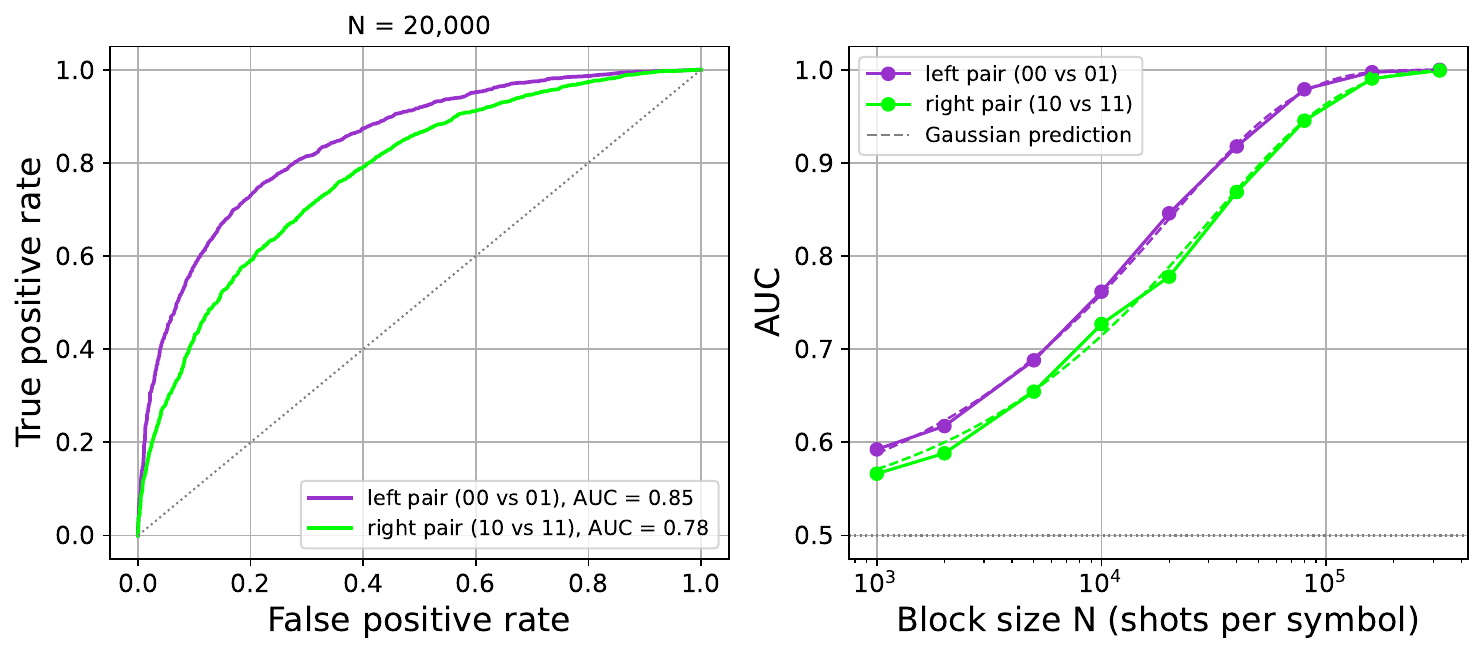}
	\caption{Left: ROC curves of the $R$-based discrimination within each pair at
	$N=\num{2e4}$; the dotted diagonal is chance level. Right: AUC as a
	function of the block size for the two pairs; the dashed lines are the
	Gaussian prediction described in the text.}
	\label{fig:roc}
\end{figure}

Figure~\ref{fig:roc} shows the ROC curves at $N=\num{2e4}$ and the AUC as a
function of the block size. The discrimination improves steadily with $N$,
following the $1/\sqrt{N}$ narrowing of the $R$ distributions: the AUC grows from
\num{0.59} (left pair) and \num{0.57} (right pair) at $N=\num{e3}$ to \num{0.92}
and \num{0.87} at $N=\num{4e4}$, and becomes compatible with one at the largest
blocks. There the fixed-threshold decoder still misreads \SI{11}{\percent} of the
symbols (the symbol error rate at $N=\num{3.2e5}$ in Table~\ref{tab:errors}):
those residual errors come from the placement of the thresholds, not
from an overlap of the $R$ distributions. The left pair discriminates
better at every $N$ because the separation of its test centroids in $R$
(\num{0.013}) is larger than that of the right pair (\num{0.011}). Like the error rates, the AUC follows from
the same Gaussian picture: $\mathrm{AUC}=\Phi\big(\Delta_R\sqrt{N}/(1.35\sqrt{2})\big)$,
with $\Delta_R$ the $R$ separation of the pair and the factor $\sqrt{2}$
accounting for the spread of the difference of two independent estimates; this
prediction, dashed in Fig.~\ref{fig:roc}, reproduces both curves over the whole
grid with nothing adjusted. This uncovers an apparent paradox: the
pair with the larger intrinsic separation is also the one that the
fixed-threshold decoder misreads most (Fig.~\ref{fig:confusion}), because the
calibration happened to place its threshold almost exactly on the 00 test
centroid. With the two halves in exchanged roles the weak state would have been
01, with a margin six times larger; the original, randomly chosen assignment of
the even and odd shots is nevertheless kept throughout, since exchanging the
halves would only hide a mismatch that any single calibration on finite data can
produce.

\subsection{Information per symbol}
\label{sec:err-info}

The overall throughput is summarized by the mutual information between the sent
symbol $X$ and the decoded symbol $Y$
\begin{equation}
	\label{eq:mutinfo}
	I(X;Y) = \sum_{x,y} p(x,y)\,\log_{2}\frac{p(x,y)}{p(x)\,p(y)}
\end{equation}
where the confusion matrix provides the conditional probabilities $p(y|x)$, each
row stating how a given sent symbol is decoded; with equal priors $p(x)=1/4$ the
joint distribution is $p(x,y)=p(y|x)/4$, and $p(y)=\sum_{x}p(x,y)$ is the
marginal distribution of the decoded symbol. It counts the bits per symbol that actually cross the channel: \num{2}
for a perfect channel, \num{0} for a useless one. It grows from \num{0.23} bits
at $N=\num{e3}$ to \num{1.25} at $N=\num{4e4}$ and \num{1.63} at $N=\num{3.2e5}$
(Table~\ref{tab:errors}). Decomposing it over the two bit
channels shows the first bit saturating at its full capacity of \num{1.00} bits,
while the second delivers \num{0.22} bits at $N=\num{4e4}$ and \num{0.55} at the
largest blocks, limited by the slow convergence discussed above.

The trade-off
between rate and reliability can also be read from the ratio $I/N$, the
information carried per shot, which decreases monotonically from \num{2.3e-4}
bits per shot at $N=\num{e3}$ to \num{5.1e-6} at $N=\num{3.2e5}$: the mutual
information grows more slowly than the block size. A fixed budget of shots thus
buys more information when split into small blocks: \num{3.2e5} shots carry
\num{1.6} bits as a single block, but about \num{74} bits as \num{320} blocks of
$N=\num{e3}$. The comparison is fair because the mutual information already discounts the
decoding errors. Written as $I(X;Y)=H(X)-H(X|Y)$, it is the entropy of the sent
symbol, $H(X)=\num{2}$ bits, minus the conditional entropy $H(X|Y)$, the
uncertainty on the sent symbol that remains once the decoded one is known. By
Shannon's noisy-channel coding theorem~\cite{shannon:communication-1948} this
difference is precisely the payload that an error-correcting code can deliver:
the code must spend $H(X|Y)$ bits per symbol of redundancy, which the receiver
uses to locate and correct the errors, and no code can do better. Large blocks
therefore pay off only when a low symbol error rate is required without coding. Extracting more
information from the same data requires a better
decoder than the fixed thresholds, which is the subject of the next section.

\section{Machine-learning discrimination}
\label{sec:ml}

The error analysis leaves an open question: does the order-zero decoder lose
information that a better receiver could recover? This section answers it with a
set of machine-learning classifiers trained to discriminate the four states,
keeping the comparison with the threshold decoder as fair as possible. Two
questions are kept separate by design: whether a better decision rule helps when
given the same information as the thresholds, and whether the blocks carry
useful information beyond $(\langle m\rangle, R)$.

\subsection{Features and training protocol}
\label{sec:ml-protocol}

Each block of $N$ shots becomes one sample described by five features: the
signal-arm mean $\langle m_2\rangle$, the noise reduction factor $R$, the
signal-arm Fano factor $F_2$, the correlation coefficient $\Gamma$ and the idler
mean $\langle m_1\rangle$. All five are available to the receiver, since the
idler counts already travel on the classical channel (Sec.~\ref{sec:decoding}).
Every classifier is trained twice: on the base set $(\langle m_2\rangle, R)$,
the same information used by the thresholds, and on the extended set with all
five features.

The data follow the protocol of Sec.~\ref{sec:err-method}. Training blocks are
drawn from the calibration pools (\num{1500} per state and block size) and test
blocks from the disjoint test pools (\num{3000} per state), with the same
bootstrap emulation; the threshold decoder is re-evaluated on the same test
blocks, so all decoders are compared on identical data, and its rates differ
from Table~\ref{tab:errors} only by the binomial fluctuations of a new
emulation. Block sizes span $N=\num{5e3}$ to $\num{8e4}$. The hyperparameters
are fixed to standard values, and the comparison is insensitive to them: a scan
around the defaults moves the error rates by less than \num{0.01}. The size of
the training set is not a limitation either: retraining on one half or one
quarter of the blocks leaves the error rates unchanged within the binomial
uncertainty (for the random forest at $N=\num{2e4}$, between \num{0.112} and
\num{0.116}), so the bottleneck is the information carried by the features, and
the calibration mismatch, rather than the size of the training set. A reduction of the features to principal
components was deliberately avoided: with five physical quantities there is no
dimensionality to reduce, and anonymous linear combinations would have hidden
the physical reading of the results below. Quoted uncertainties are binomial
standard errors over the \num{12000} test blocks.

\subsection{The classifiers}
\label{sec:ml-roster}

Five supervised classifiers span the main model families:
\begin{description}
	\item[$k$-nearest neighbors (KNN)] looks up the $k=25$ training blocks
	closest to the one being decoded, with distances measured on standardized
	features, and takes the most common label among them; all neighbors count
	equally. Since it assumes nothing about the shape of the classes, it serves
	as the reference against which the model-based classifiers are read.
	\item[Support vector machine (SVM)] separates the classes with the surface
	that keeps the largest possible margin from the training blocks on either
	side; a Gaussian kernel lets the surface bend where the data require it.
	With few features and well-populated classes, as here, it is usually the
	strongest general-purpose discriminator.
	\item[Random forest (RF)] grows $200$ decision trees, each on a different
	resampling of the training set, and takes their majority vote. Averaging
	many weakly correlated trees gives a stable classifier and, as a by-product,
	the feature importances: the fraction of the decisions attributable to each
	feature.
	\item[Extreme gradient boosting (XGBoost)] also combines decision trees, but
	sequentially, with every new tree fitted to the errors left by the previous
	ones; it is the standard reference for this kind of tabular data.
	\item[Quadratic discriminant analysis (QDA)] models each class as a Gaussian
	with its own mean and covariance and assigns a block to the most probable
	class. This amounts to minimizing the same Mahalanobis distance that defines
	the confidence ellipses of Sec.~\ref{sec:constellation}: QDA is the decision
	rule implied by the Gaussian picture used throughout this report, and the
	optimal one if that picture is exact.
\end{description}
A sixth, unsupervised model is considered in Sec.~\ref{sec:ml-gmm}.

\subsection{Results}
\label{sec:ml-results}

\begin{figure}[htp]
	\centering
	\includegraphics[width=\textwidth]{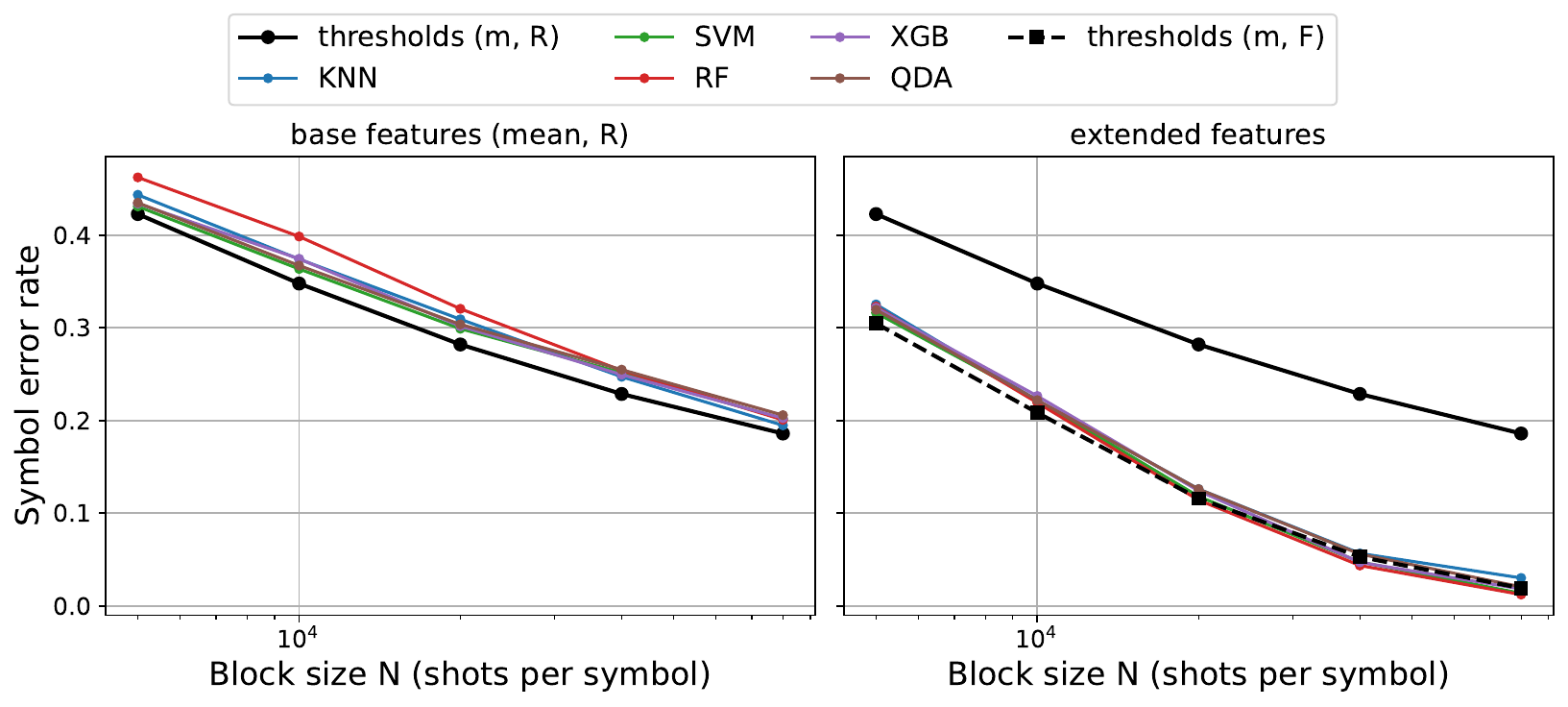}
	\caption{Symbol error rate of the classifiers as a function of the block
	size, on the base feature set (left) and on the extended one (right); the
	solid black curve is the $(\langle m\rangle, R)$ threshold decoder, evaluated
	on the same test blocks, and the dashed black curve in the right panel is the
	$(\langle m\rangle, F)$ threshold decoder of Sec.~\ref{sec:ml-fano}.}
	\label{fig:mlser}
\end{figure}

Table~\ref{tab:ml} and Fig.~\ref{fig:mlser} collect the symbol error rates. On
the base set no classifier beats the thresholds: all five sit at or slightly
above the threshold rates at every block size (at $N=\num{4e4}$, between
\num{0.247} and \num{0.255} against \num{0.229}). This includes QDA, which
would be optimal for Gaussian clouds: the limitation of the physical decoder is
not the shape of its decision boundaries but the information in
$(\langle m\rangle, R)$, and the slight degradation shows the classifiers
adapting to the calibration realization, the same mismatch that limits the
thresholds.

On the extended set the picture changes completely: the error rate drops to
\num{0.044} at $N=\num{4e4}$ and \num{0.012} at $N=\num{8e4}$ (random forest), a
factor $5$ to $15$ below the $(\langle m\rangle, R)$ thresholds, with all five
models within a factor of two to three of each other. The gain therefore comes from the added information; which of the five
rules is used matters little. The feature importances say where that
information lives: at $N=\num{2e4}$ the mean carries about \num{0.43} of the
decisions, the Fano factor $F_2$ \numrange{0.36}{0.44}, the noise reduction
factor only \numrange{0.07}{0.13}, and $\Gamma$ and $\langle m_1\rangle$ are
negligible. The second bit, the number of modes of the noise, is read far
better from the variance of the signal arm alone than from the joint quantity
$R$: the same conclusion suggested by the additivity check of
Sec.~\ref{sec:additivity}, where the four states differ in $F$ by up to
\num{0.04}.

\begin{table}[htp]
	\centering
	\sisetup{separate-uncertainty=false}
	\caption{Symbol error rate of all decoders as a function of the block size,
	on the same test blocks. The first two rows are the threshold decoders built
	on $(\langle m\rangle, R)$ (Sec.~\ref{sec:decoding}) and on
	$(\langle m\rangle, F)$ (Sec.~\ref{sec:ml-fano}); the classifiers appear with
	the base and the extended feature set. Uncertainties are binomial standard
	errors over the \num{12000} test blocks, in compact form.}
	\label{tab:ml}
	\small
	\setlength{\tabcolsep}{4.5pt}
	\begin{tabular}{l S[table-format=1.4(2)] S[table-format=1.4(2)] S[table-format=1.4(2)] S[table-format=1.4(2)] S[table-format=1.4(2)]}
		\toprule
		 & \multicolumn{5}{c}{block size $N$} \\
		\cmidrule(lr){2-6}
		{decoder} & {\num{5000}} & {\num{10000}} & {\num{20000}} & {\num{40000}} & {\num{80000}} \\
		\midrule
		thresholds $(\langle m\rangle, R)$ & 0.4228(45) & 0.3478(43) & 0.2820(41) & 0.2286(38) & 0.1859(36) \\
		thresholds $(\langle m\rangle, F)$ & 0.3048(42) & 0.2083(37) & 0.1158(29) & 0.0528(20) & 0.0191(12) \\
		\midrule
		KNN, base & 0.4435(45) & 0.3741(44) & 0.3092(42) & 0.2472(39) & 0.1948(36) \\
		KNN, extended & 0.3252(43) & 0.2218(38) & 0.1260(30) & 0.0566(21) & 0.0301(16) \\
		SVM, base & 0.4309(45) & 0.3635(44) & 0.2990(42) & 0.2516(40) & 0.2018(37) \\
		SVM, extended & 0.3162(42) & 0.2218(38) & 0.1180(29) & 0.0473(19) & 0.0135(11) \\
		RF, base & 0.4623(46) & 0.3985(45) & 0.3205(43) & 0.2539(40) & 0.2007(37) \\
		RF, extended & 0.3229(43) & 0.2192(38) & 0.1141(29) & 0.0435(19) & 0.0122(10) \\
		XGBoost, base & 0.4341(45) & 0.3744(44) & 0.3018(42) & 0.2494(39) & 0.2025(37) \\
		XGBoost, extended & 0.3216(43) & 0.2264(38) & 0.1241(30) & 0.0470(19) & 0.0182(12) \\
		QDA, base & 0.4349(45) & 0.3672(44) & 0.3037(42) & 0.2548(40) & 0.2057(37) \\
		QDA, extended & 0.3195(43) & 0.2219(38) & 0.1259(30) & 0.0558(21) & 0.0205(13) \\
		\bottomrule
	\end{tabular}
\end{table}

\subsection{A Fano-based threshold decoder}
\label{sec:ml-fano}

If the discrimination information is carried by the Fano factor, machine learning
should not be needed to collect it. The order-zero scheme of
Sec.~\ref{sec:decoding} is therefore rebuilt with $F_2$ in place of $R$: the
first bit from the mean threshold, the second from a Fano threshold within each
half ($F_{\mathrm{th}}^{L}=\num{1.088}$, $F_{\mathrm{th}}^{R}=\num{1.097}$,
midpoints of the calibration values; the direction is uniform, since the
few-mode noise of states 01 and 11 gives the larger $F$). This purely physical
decoder nearly matches the best classifier at every block size
(Table~\ref{tab:ml}), and beats it at the smallest ones.

\begin{figure}[htp]
	\centering
	\includegraphics[width=.9\textwidth]{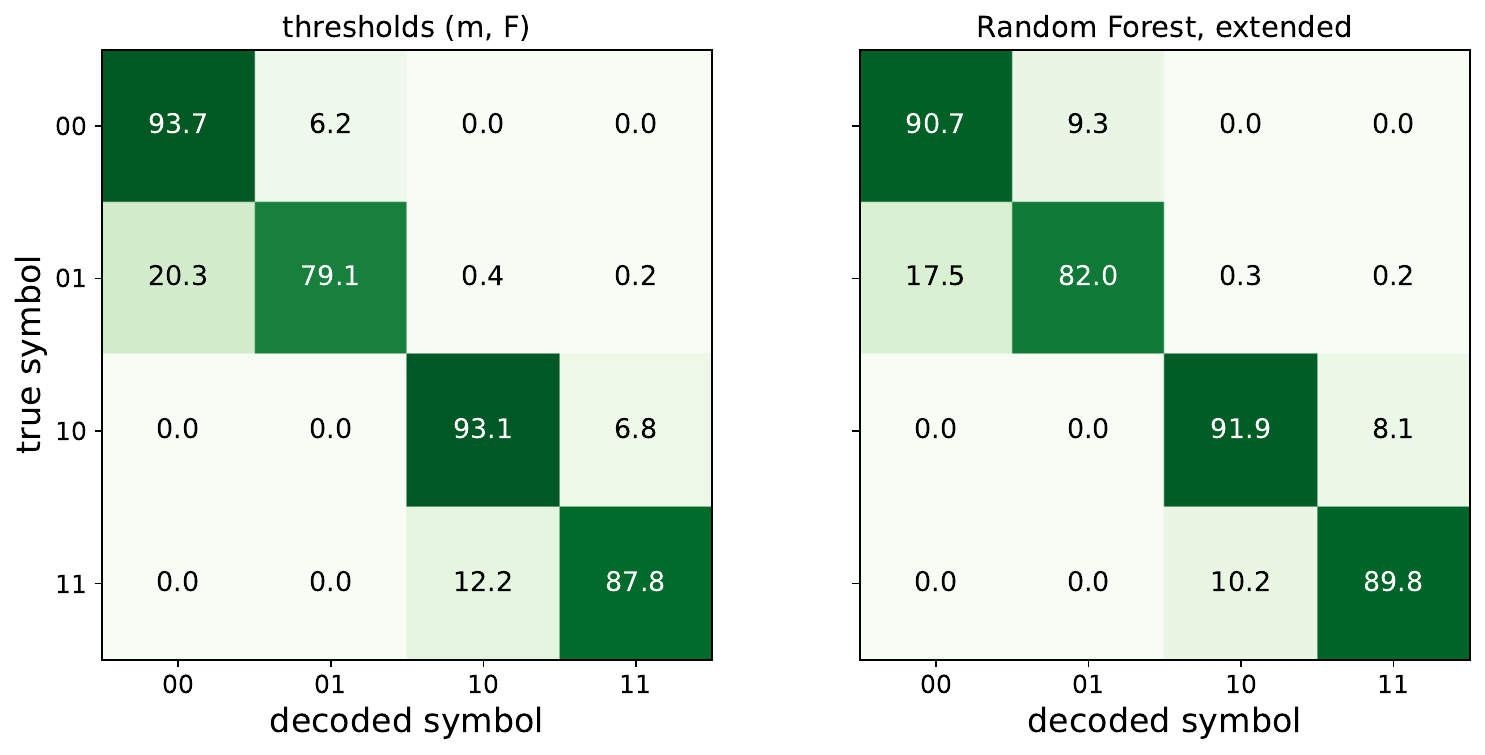}
	\caption{Confusion matrices at $N=\num{2e4}$ (rows: sent symbol; columns:
	decoded symbol; entries in percent) for the $(\langle m\rangle, F)$ threshold
	decoder (left) and the random forest on the extended features (right).
	Compare with Fig.~\ref{fig:confusion} at the same block size: the errors
	remain within the pairs but are now symmetric, with no state left on a
	threshold.}
	\label{fig:mlconf}
\end{figure}

The margin machinery of Sec.~\ref{sec:err-vs-N} explains why. The Fano factor is
itself a variance over a mean, so its estimates spread as
$\sigma(F)\simeq F\sqrt{2/N}$; the measured $\sigma(F)\sqrt{N}\approx\num{1.56}$
is constant over the whole grid. The test centroids sit
\numrange{0.009}{0.017} away from the calibrated Fano thresholds: the worst
margin is fourteen times the worst margin in $R$ (\num{0.0006},
Sec.~\ref{sec:err-vs-N}), which is the whole story of the improvement. The same
margin prediction used for Fig.~\ref{fig:errN} reproduces the measured rates of
this decoder over the whole grid, and no state is left on a threshold: the
confusion matrix, shown in Fig.~\ref{fig:mlconf} next to that of the best
classifier, is symmetric within the pairs, without the coin-flip state of
Fig.~\ref{fig:confusion}.

\begin{figure}[htp]
	\centering
	\includegraphics[width=\textwidth]{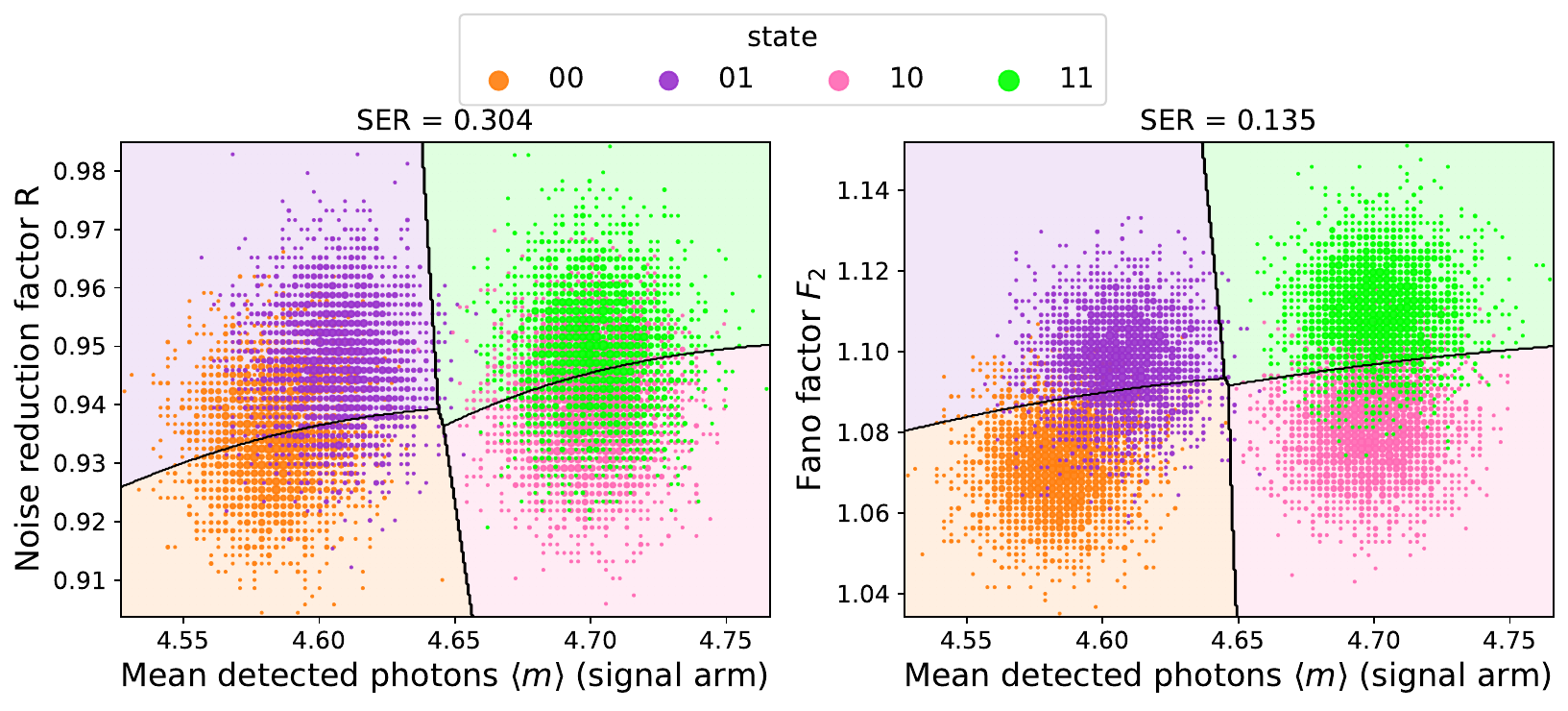}
	\caption{Decision regions learned by QDA at $N=\num{2e4}$ when trained on
	$(\langle m\rangle, R)$ (left) and on $(\langle m\rangle, F)$ (right), with
	the test clouds overlaid; the panel titles report the test symbol error
	rates. On the left the learned partition is close to the threshold map of
	Fig.~\ref{fig:protocol}; on the right the states are separated far better,
	with only residual overlap within the pairs.}
	\label{fig:mlregions}
\end{figure}

The decision regions make this picture visible. Figure~\ref{fig:mlregions}
shows them for a single classifier, taken as a representative example: QDA,
chosen because its regions have a direct physical reading, being the Bayes rule
of the Gaussian clouds used throughout this report. Trained on
$(\langle m\rangle, R)$ it learns a partition close to the threshold map of
Fig.~\ref{fig:protocol}, with curved boundaries in place of straight ones but
the same structure. Trained on $(\langle m\rangle, F)$ it separates the four
states far better, though the clouds within each pair still overlap, which is the
residual error of the Fano decoder above.

One property of this decoder must be stated clearly. $F_2$ is a classical,
single-arm quantity: the two bits are decoded from the signal arm alone, without
Alice's idler, but for the same reason no quantum quantity protects the link any
longer. An eavesdropper who reproduces mean and Fano factor of the states, which
is classically possible, is invisible to this decoder: with nothing quantum left
in the decoding, the scheme is ordinary classical communication. What
restores the role of the twin beam is the noise reduction factor, the only
quantity that an intercept-and-resend attack cannot counterfeit:
Sec.~\ref{sec:ml-revised} rebuilds the protocol around this division of labor,
with $(\langle m\rangle, F)$ carrying the data and $R$ certifying the channel.

\subsection{Blind discovery of the alphabet}
\label{sec:ml-gmm}

The last model asks a different question: is the alphabet structure discoverable
without labels? A Gaussian mixture model (GMM) fits the blocks with Gaussian
components by expectation maximization, never seeing the true symbols. The number
of components is chosen by the Bayesian information criterion, which scores a fit
by its likelihood but adds a penalty growing with the number of parameters, so
that a further component is preferred only when the data justify its cost; this
avoids the trivial preference for ever more components. Left to this criterion,
and blind to the four intended states, it counts the statistically distinct
clusters in the data: three in the $(\langle m\rangle, R)$ plane, four in the
$(\langle m\rangle, F)$ plane and on the full feature set.

Whether these clusters are the states is a separate question, answered by
matching each fitted component to the true symbol of the blocks it contains and
measuring how often the two agree. In $(\langle m\rangle, F)$ and on the extended
features the four components fall one per state, and the match is correct
\SI{91}{\percent} and \SI{92}{\percent} of the time: the blind fit recovers the
alphabet. In the $(\langle m\rangle, R)$ plane only three clusters exist, so the
four states cannot be matched one to one; the right pair, whose clouds overlap
(Fig.~\ref{fig:constellation}), is a single cluster and the agreement is capped
at \SI{64}{\percent}. Because the criterion chose the number of components on its own, the recovered
structure is a genuine feature of the data. The extended-feature agreement is even
higher than the accuracy of the supervised classifiers at the same block size,
because the mixture is fitted on the received stream itself and is immune to the
calibration mismatch: unsupervised recalibration is a practical remedy against
slow drifts.

The blind discovery carries a security implication. The four-state structure is
recovered from $(\langle m\rangle, F)$ alone, both quantities of the signal arm,
so an eavesdropper who listens only to the transmitted beam can learn the
alphabet and read the message. The protocol therefore protects the link by
revealing an intrusion, not by keeping the content secret, which is the role of
the noise reduction factor examined in the next section.

\subsection{A revised protocol}
\label{sec:ml-revised}

The results of this section suggest a natural redesign of the protocol, with two
separate planes. The data plane decodes both bits from
$(\langle m\rangle, F)$, on the signal arm alone, with the error rates of
Table~\ref{tab:ml}. The control plane monitors the noise reduction factor, whose
sub-shot-noise value cannot be reproduced by any classical resend and therefore
remains the only guarantee against an intercept-and-resend attack. Since each
block is decoded before the check, the monitor can compare $R$ with the
reference value of the identified state, a sharper test than a global threshold,
and Alice needs to disclose her idler counts only for the blocks used in the
check, which can be a random sample unknown to the eavesdropper in advance.

The two planes define a range of operating points. Sending both bits through
$(\langle m\rangle, F)$ uses the full rate, at the residual second-bit error of
Table~\ref{tab:ml}; dropping the fragile second bit and carrying a single bit
through the mean alone halves the rate but is essentially error-free (a
first-bit error of \num{0.003} at $N=\num{2e4}$, negligible beyond). This
conservative extreme, one classical bit read from the mean with $R$ as the sole
security check, is the binary protocol of Ref.~\cite{razzoli:hybrid-2025}, which
the four-state scheme thus contains as a limiting case.

The response of this monitor to an attack is the subject of the next section.

\section{Detection of an intercept-and-resend attack}
\label{sec:attack}

In the revised protocol of Sec.~\ref{sec:ml-revised} the two bits travel on
classical quantities of the signal arm, and the noise reduction factor is kept
aside as a channel monitor. This section puts that monitor to the test: an
intercept-and-resend attack is emulated on the data, the response of $R$ and
$F$ is measured as a function of the attacked fraction, and the probability of
detecting the intrusion is quantified with the same statistical tools used for
the error analysis.

\subsection{Attack emulation}
\label{sec:attack-model}

The eavesdropper, conventionally called Eve, taps the signal arm between Alice
and Bob. Of every block of
$N$ shots she intercepts a fraction $f$, measures it, and resends light with
the same mean value, so the mean, and with it the first bit, is unchanged.
Eve is taken to be ideal, estimating the intercepted mean without error; a real
detector of finite efficiency would also perturb the mean, adding a classical
signature on top of the rise of $R$, so the ideal case treated here is the
hardest to detect.
Alice's idler never leaves her side: Bob computes $R$ by pairing his possibly
attacked signal counts with clean idler counts, so on the attacked shots the
twin-beam correlation is replaced by no correlation at all and $R$ has to
rise. Two eavesdroppers of increasing skill are considered.
The first, called the \textit{Poisson resend}, is the attack considered in
Ref.~\cite{razzoli:hybrid-2025}: Eve resends coherent-like light
at the mean she measured. The mean is right, but the resent variance is the
Poissonian one, well below the super-Poissonian variance of the states, so the
Fano factor of the block is pulled toward $1$. The second, called the
\textit{record-and-resend}, resends the very photon numbers she recorded, in a
different order (equivalently, she prepares pseudo-thermal light with exactly
the measured statistics). The multiset of counts is identical, so mean,
variance and Fano factor of the block are preserved exactly, and only the
shot-by-shot pairing with the idler is destroyed. This is the strongest
attack available to an eavesdropper without access to the idler: no
single-arm quantity can distinguish the resent light from the genuine one.
This second attack is not considered in Ref.~\cite{razzoli:hybrid-2025},
whose decoding never uses the signal-arm Fano factor; it becomes the relevant
threat here because the revised protocol reads the second bit from the Fano
factor, which the Poisson resend visibly perturbs.

The attack is emulated on the bootstrap blocks of the test pools, with
thresholds and references calibrated on the disjoint calibration pools, as in
Sec.~\ref{sec:err-method}: in each block a random fraction $f$ of the
signal-arm counts is replaced by Eve's resend. The expected
response follows from the covariance $C$ of the two arms. Replacing a
fraction $f$ of the shots with uncorrelated light of the same statistics
scales the covariance to $(1-f)\,C$, so the variance of the difference grows
by $2fC$ and the noise reduction factor of Eq.~\eqref{eq:R} rises linearly
\begin{equation}
	\label{eq:attack-R}
	R(f) = R_0 + f\,\frac{2C}{\langle m_1\rangle + \langle m_2\rangle}
\end{equation}
where $R_0$ is the unattacked value. With the measured covariances the slope
is between \num{0.130} and \num{0.133} for the four states. The Poisson
resend also lowers the variance of the signal arm, which removes part of the
rise, giving a smaller slope of about \num{0.09}, and pulls the Fano factor
down linearly, $F(f) = 1 + (1-f)(F_2 - 1)$. At $f=1$ the record-and-resend
must land on
$R = (\sigma_1^{2}+\sigma_2^{2})/(\langle m_1\rangle+\langle m_2\rangle)$,
the mean-weighted average of the two single-arm Fano factors, since all
correlation is gone while the marginal statistics are intact: a parameter-free
sanity check of the emulation. The idler carries no added noise, so this
endpoint lies below $F_2$.

\subsection{Response of the noise reduction factor}
\label{sec:attack-R}

\begin{figure}[htp]
	\centering
	\includegraphics[width=\textwidth]{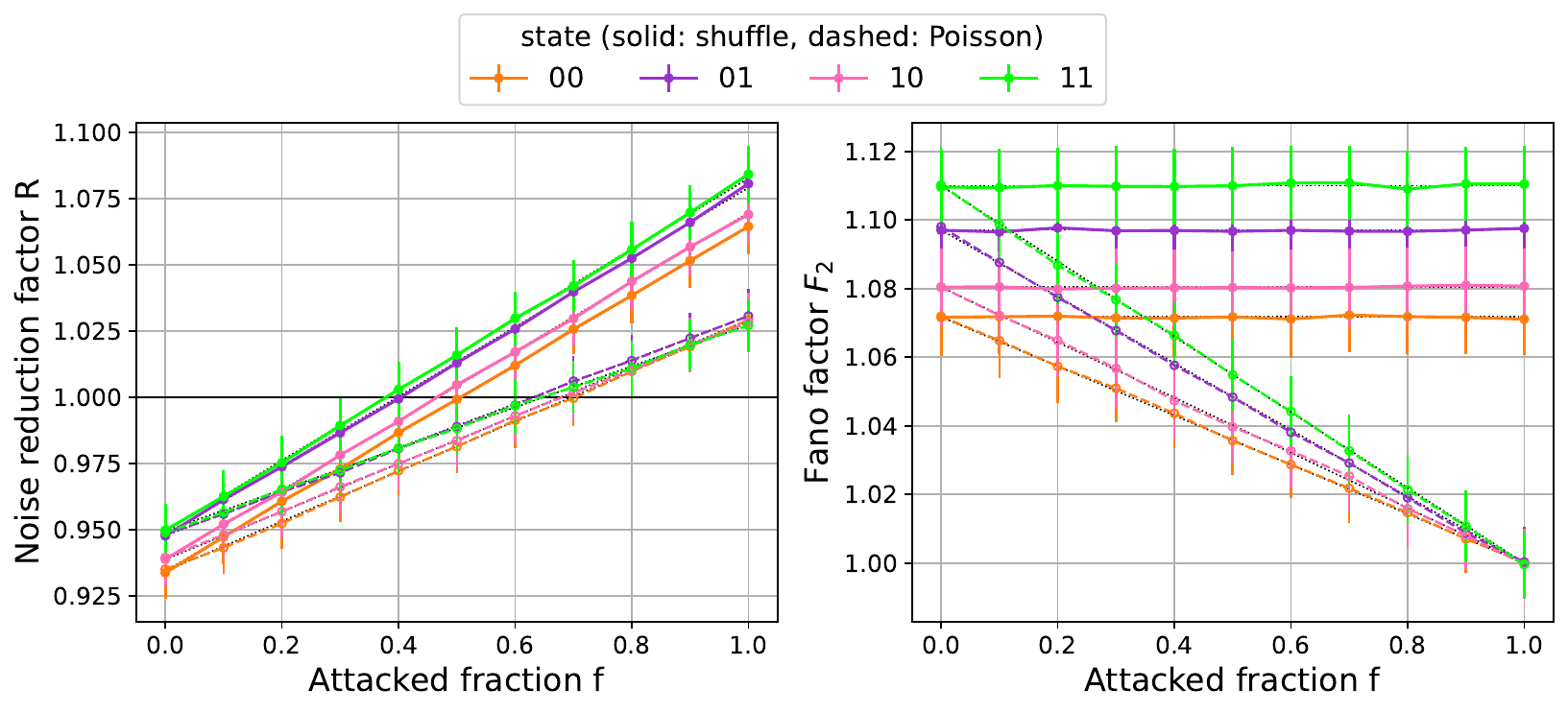}
	\caption{Response of the noise reduction factor (left) and of the
	signal-arm Fano factor (right) to the attacked fraction $f$, for the four
	states at $N=\num{2e4}$. Filled markers with solid lines: record-and-resend
	attack; open markers with dashed lines: Poisson resend; dotted black lines:
	predictions built from the measured covariances, with no adjusted
	parameters (Eq.~\eqref{eq:attack-R} and Sec.~\ref{sec:attack-model}). Error
	bars are the block-to-block spread. The record-and-resend leaves the Fano
	factor exactly unchanged, while the Poisson resend pulls it toward $1$.}
	\label{fig:attackresp}
\end{figure}

Figure~\ref{fig:attackresp} shows the response at $N=\num{2e4}$. The noise
reduction factor rises linearly with $f$ for all four states and both
attacks, and the predictions built from the measured covariances, with
nothing adjusted, reproduce the measured values to the third decimal digit
over most of the range (for state 00 under record-and-resend, \num{0.9992}
measured against \num{0.9993} predicted at $f=\num{0.5}$) and the $f=1$
endpoint within \num{0.002}. The Fano factor separates the two attacks: under
the record-and-resend it stays exactly at the unattacked value, while under
the Poisson resend it drops linearly toward $1$.

\begin{figure}[htp]
	\centering
	\includegraphics[width=.8\textwidth]{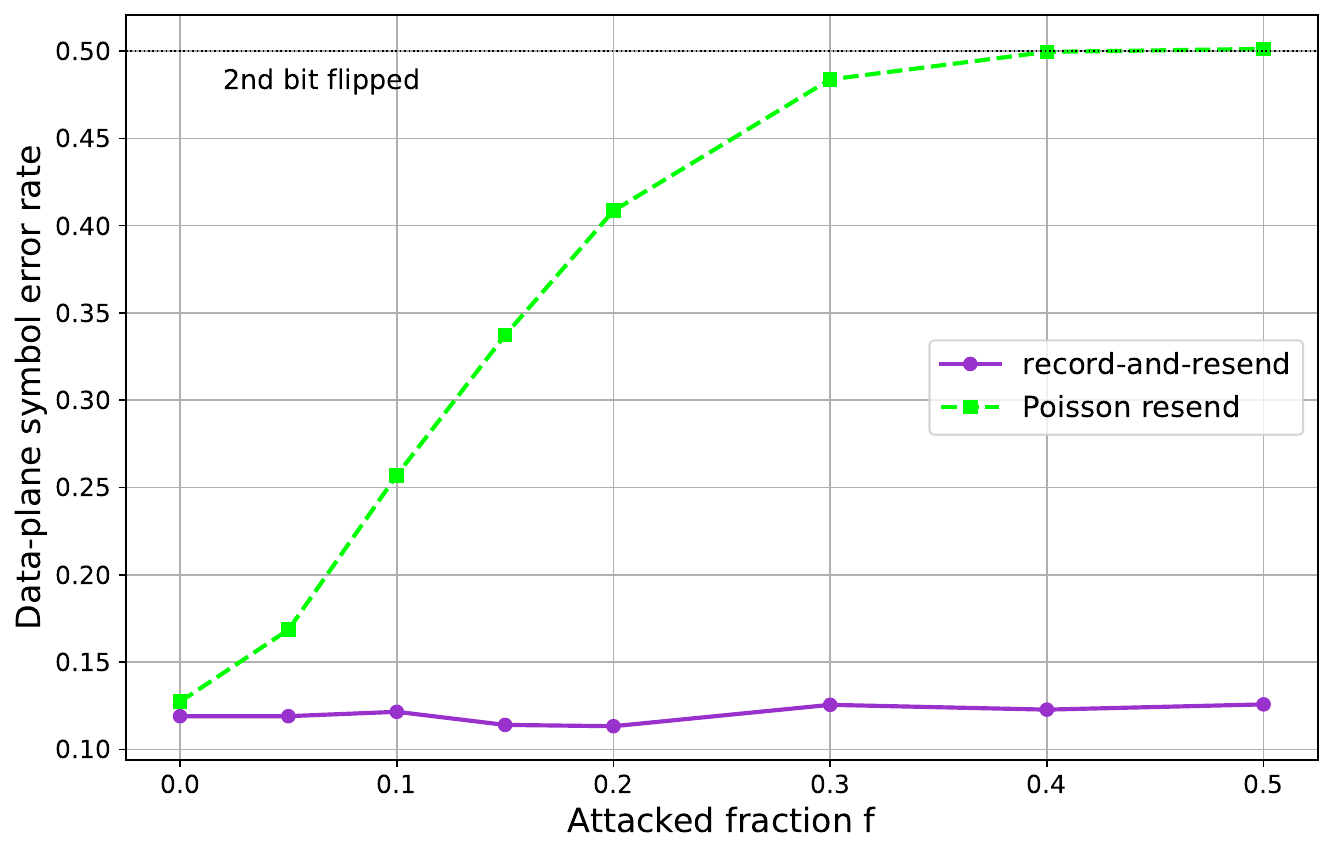}
	\caption{Symbol error rate of the $(\langle m\rangle, F)$ data-plane decoder
	as a function of the attacked fraction $f$, at $N=\num{2e4}$. The
	record-and-resend leaves the decoder untouched (flat rate), while the Poisson
	resend drives it toward \num{0.5}, a random second bit, as the collapsing Fano
	factor crosses the decision thresholds. Only the noise reduction factor sees
	the record-and-resend (Fig.~\ref{fig:attackdet}).}
	\label{fig:attackser}
\end{figure}

The consequences for the data plane are opposite (Fig.~\ref{fig:attackser}).
Under the record-and-resend the $(\langle m\rangle, F)$ decoder is untouched: its symbol error rate stays
at the unattacked value (between \num{0.11} and \num{0.13} at $N=\num{2e4}$)
up to $f=\num{0.5}$, with no trend. Eve reads the message and
leaves no classical trace. Under the Poisson resend, instead, the collapsing
Fano factor drags the blocks across the $F$ thresholds: the symbol error
rate grows from \num{0.13} to \num{0.41} already at $f=\num{0.2}$ and
saturates at \num{0.5}. That value is not the sign of a random second bit: the
Fano factor of every block ends up below its threshold, so the two states that
carry that bit as $1$ are always misread while the other two are always right,
which averages to one half. This attack
announces itself in the data plane before any security check, which
confirms the division of labor found in Sec.~\ref{sec:ml-fano}: the Fano
factor is a useful classical alarm, but one the eavesdropper can falsify
exactly, and the noise reduction factor is the only quantity that neither
attack can counterfeit.

A last observation on Fig.~\ref{fig:attackresp}: $R$ crosses the classical
boundary $R=1$ only for $f\gtrsim\num{0.4}$. Losing the nonclassicality
certificate is therefore a very insensitive symptom; the detection below
works on the deviation of $R$ from its per-state reference, which responds
to much smaller attacked fractions.

\subsection{Statistical detection}
\label{sec:attack-detection}

The monitor follows the pipeline of the revised protocol. For every block Bob
first decodes the state from $(\langle m\rangle, F)$, then computes $R$ with
Alice's idler counts and compares it with the calibrated reference of the
identified state through the one-sided statistic
\begin{equation}
	\label{eq:attack-z}
	z = \frac{\hat R - R_{\mathrm{ref}}}{\sigma(R)}
	\qquad
	\sigma(R) = \frac{1.35}{\sqrt{N}}
\end{equation}
with $\sigma(R)$ the spread measured in Sec.~\ref{sec:err-vs-N}, verified to
hold for the attacked blocks as well. The alarm is raised when
$z>\num{1.645}$, the \SI{95}{\percent} quantile of the standard Gaussian, so
that a clean block with a perfect reference triggers a false alarm with the
nominal probability of \SI{5}{\percent}. The test is one-sided because the
attack can only raise $R$.

\begin{figure}[htp]
	\centering
	\includegraphics[width=\textwidth]{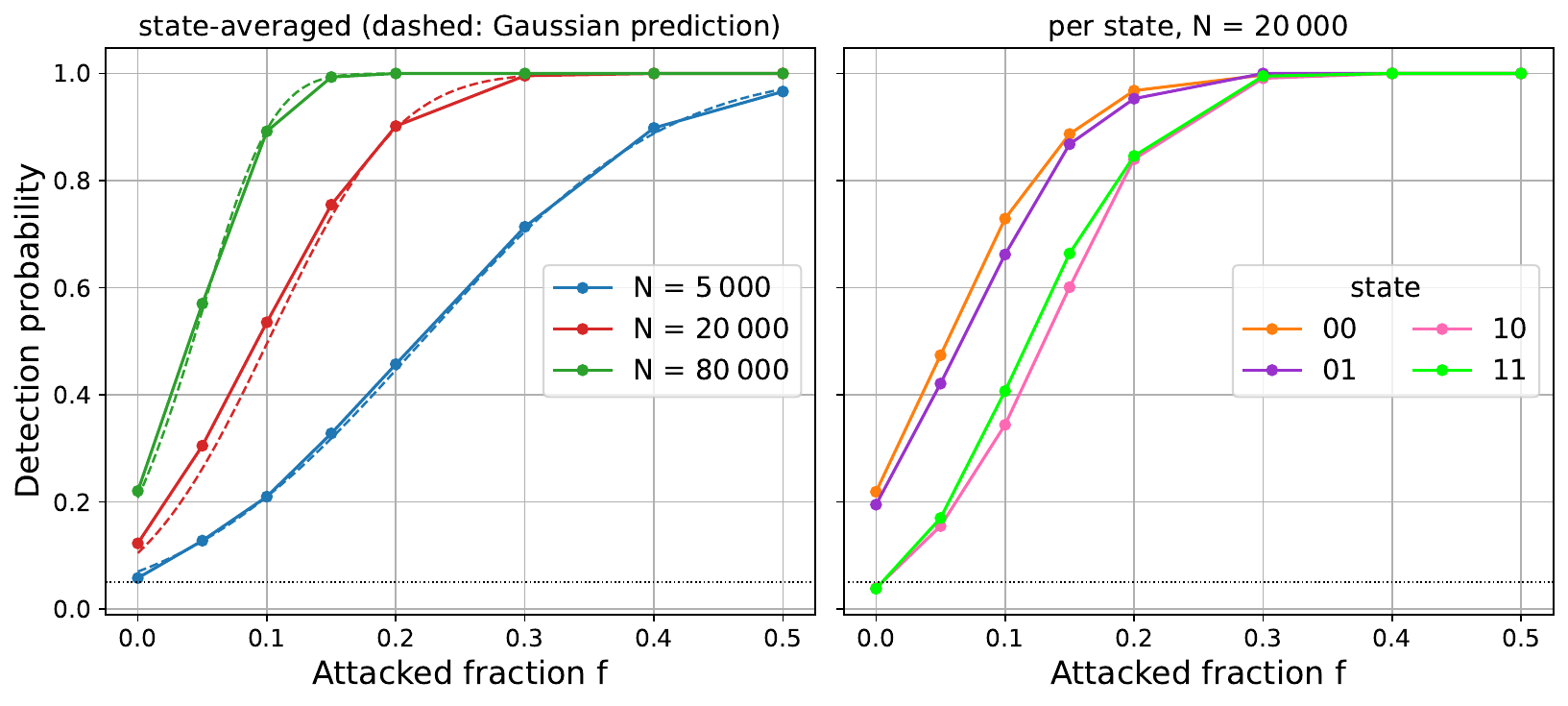}
	\caption{Probability of detecting the record-and-resend attack as a
	function of the attacked fraction $f$. Left: average over the four states
	at the three block sizes; the dashed lines are the Gaussian prediction
	described in the text, with no adjusted parameters. Right: per-state
	detection probability at $N=\num{2e4}$. The dotted horizontal line marks
	the nominal \SI{5}{\percent} false-alarm level; at $f=0$ the curves start
	above it, mostly because of the calibration mismatch of the references
	(largest for state 00).}
	\label{fig:attackdet}
\end{figure}

Figure~\ref{fig:attackdet} shows the measured detection probability against
the record-and-resend attack, the one invisible to everything else. The
power grows with $f$ and with the block size: a \SI{90}{\percent} detection
probability is reached at $f\approx\num{0.40}$, \num{0.20} and \num{0.10}
for $N=\num{5e3}$, \num{2e4} and \num{8e4}, the $1/\sqrt{N}$ improvement
expected from Eq.~\eqref{eq:attack-z}. The whole family of curves is
reproduced, with nothing adjusted, by the Gaussian expression
$\Phi\big((\Delta + s f)\sqrt{N}/1.35 - 1.645\big)$, where $s$ is the slope
of Eq.~\eqref{eq:attack-R} and $\Delta$ is the offset between the test
centroid and the calibrated reference of the state. The Poisson resend is
detected even faster (an alarm probability of \num{0.81} at $f=\num{0.2}$
and $N=\num{2e4}$), partly because the broken decoding compares many of its
blocks against the reference of the wrong state; for that attack the monitor
is anyway redundant, since the data plane itself fails visibly.

The offsets $\Delta$ deserve the same honesty applied in
Sec.~\ref{sec:err-vs-N}. At $f=0$ the measured false-alarm probability is
\num{0.058}, \num{0.123} and \num{0.220} at the three block sizes, above the
nominal \SI{5}{\percent}. The excess has two sources, separated by rerunning
the monitor with different references. The dominant one is the calibration
mismatch: the $R$ centroid of the test half of state 00 sits \num{0.0091}
above its calibrated reference, about twice the sampling error of the
reference itself, the same fluctuation that produced the decoding floor of
Sec.~\ref{sec:err-vs-N}. With oracle references, the test centroids
themselves, the false alarm returns to \num{0.043}, \num{0.045} and
\num{0.055}. The second, smaller source is misdecoding: at $N=\num{2e4}$
about \SI{12}{\percent} of the clean blocks are decoded as the wrong state
of their pair and tested against a reference that does not belong to them.
The sensitivity of the monitor is therefore limited by the quality of the
calibration rather than by the physics: a practical implementation needs
reference values calibrated on samples large enough, or refreshed
periodically, exactly as concluded for the decoder thresholds.

Two design features complete the picture. Since the monitor works per block
and per state, Alice needs to disclose her idler counts only for the blocks
chosen for verification, which can be a random sample unknown to Eve in
advance. And as anticipated in Sec.~\ref{sec:ml-gmm}, the
protection is detection, not secrecy: under the record-and-resend, Eve
does read the intercepted symbols, but she cannot avoid leaving her
signature in $R$, and the legitimate parties learn that the channel is
compromised.

\section{Conclusions}
\label{sec:conclusions}

The binary hybrid protocol of Razzoli et al.~\cite{razzoli:hybrid-2025} was
extended to a four-symbol alphabet, carrying two bits per transmission. The twin
beam was first characterized alone: both arms follow multi-mode thermal
statistics, the noise reduction factor stays below one at every intensity, and
fitting $R$ against the mean detected photons gives a detection efficiency
$\eta=\SI{9.0(3)}{\percent}$, close to the value reported for the reference
channel. A weak thermal noise, prepared with two mean values and two mode
numbers, was then added to the signal arm to place the four states at distinct
points of the $(\langle m\rangle, R)$ plane. Unlike the reference, where the noise
and the twin beam are merged in post-processing, here the two beams were combined
optically in the same collection fiber, one step closer to a real link.

The mean value carries the first bit and is read with negligible error, while the
noise reduction factor carries the second and separates the states only weakly,
their clouds overlapping in $R$. An order-zero threshold decoder recovers the
message, and its error rate was measured against the block size on disjoint
calibration and test data: the first bit becomes error-free at moderate block
sizes, while the second decays slowly, limited by the mismatch between the
calibrated thresholds and the test statistics rather than by counting noise. The block
size trades rate against reliability: larger blocks are more reliable, but a fixed
number of shots carries more information when split into smaller blocks, since the
mutual information grows more slowly than the block size. A
comparison with a set of machine-learning classifiers located the discriminating
information: within each pair it is carried mostly by the single-arm Fano factor,
not by $R$. With that information the classifiers cut the error rate by a factor
of \numrange{5}{15}, and a purely classical decoder on $(\langle m\rangle, F)$,
which uses the signal arm alone, nearly matches the best of them. This points to
a protocol on two planes, with
$(\langle m\rangle, F)$ carrying the data and $R$ kept aside as a per-state
monitor of the channel. Dropping the fragile second bit, so that a single
classical bit is read from the mean with $R$ as the only security check, recovers
the binary protocol of the reference as a limiting case.

The monitor was tested against an intercept-and-resend attack. Replacing a
fraction of the signal-arm shots with uncorrelated light raises $R$ linearly, an
effect reproduced by a parameter-free model, and a resend that also matches the
Fano factor leaves every single-arm quantity untouched, so that only $R$ reacts.
A per-state test on $R$ detects such an attack once it reaches about
\SIrange{10}{20}{\percent} of the shots at the larger block sizes considered, with a
sensitivity set by how well the reference values are calibrated. The nonclassical
correlation of the twin beam thus protects the link not by keeping the message
secret, which a blind clustering already reads from the signal arm, but by
exposing the intrusion.

Some extensions remain open. The alphabet can be enlarged to eight states, three
bits per transmission, by adding a third quantity and arranging the states on a
$2\times2\times2$ grid instead of the present $2\times2$. The fragile second bit
is the natural target for an error-correcting code; since a symbol is a resizable
block of the same stream, and one estimate on a longer block beats the majority
vote of several shorter ones, such a code should act on that bit with
soft-decision decoding suited to bit error rates of \numrange{0.1}{0.4}, rather
than simply enlarging the blocks. A faster acquisition would lift the main
experimental limitation of this work: with many more shots stored per state,
calibration and communication could draw on disjoint pools large enough for the
blocks to be independent without resampling, so a message would be transmitted
rather than emulated by bootstrap, and the error rate at large block sizes would no
longer be limited by the fluctuation of a single acquisition. Moving the source to telecom wavelengths, with
the detection kept in the visible through sum-frequency generation, would take the
scheme toward a deployable channel.

\cleardoublepage
\printbibliography

@article{razzoli:hybrid-2025,
	author		= {Razzoli, Luca and Pozzoli, Alex and Allevi, Alessia},
	title		= {Hybrid discrimination strategy in quantum communication based on photon-number-resolving detectors and mesoscopic twin-beam states},
	journaltitle	= {Quantum Science and Technology},
	date		= {2025},
	volume		= {10},
	number		= {4},
	pages		= {045036},
	doi		= {10.1088/2058-9565/ae05c3},
}

@article{allevi:multimode-2022,
	author		= {Allevi, Alessia and Bondani, Maria},
	title		= {Multi-mode twin-beam states in the mesoscopic intensity domain},
	journaltitle	= {Physics Letters A},
	date		= {2022},
	volume		= {423},
	pages		= {127828},
}

@article{shannon:communication-1948,
	author		= {Shannon, Claude E.},
	title		= {A Mathematical Theory of Communication},
	journaltitle	= {Bell System Technical Journal},
	date		= {1948},
	volume		= {27},
	pages		= {379-423},
	doi		= {10.1002/j.1538-7305.1948.tb01338.x},
}

@article{bondani:self-consistent-2009,
	author		= {Bondani, Maria and Allevi, Alessia and Agliati, Andrea and Andreoni, Alessandra},
	title		= {Self-consistent characterization of light statistics},
	journaltitle	= {Journal of Modern Optics},
	date		= {2009},
	volume		= {56},
	pages		= {226-231},
}

@online{hamamatsu:s15639-datasheet,
	author		= {{Hamamatsu Photonics K.K.}},
	title		= {MPPC S15639-1325PS, datasheet},
	year		= {2022},
	url		= {https://www.hamamatsu.com/us/en/product/optical-sensors/mppc/mppc_mppc-array/S15639-1325PS.html},
	urldate		= {2026-07-21},
}
\end{document}